\documentclass[submitting]{nst}

\usepackage{amsmath}
\usepackage{amssymb}
\usepackage[english]{babel}
\usepackage{bm}
\usepackage{booktabs}
\usepackage{cleveref}
\usepackage{dcolumn}
\usepackage[T1]{fontenc}
\usepackage{microtype}
\usepackage{multirow}
\usepackage{subfigure}
\usepackage{threeparttable}

\graphicspath{{figures/}}
\begin{document}
\nolinenumbers

\title{Proton-Number Fluctuations in $^{\text{238}}$U+$^{\text{238}}$U Collisions at $\sqrt{\textit{s}_{\text{NN}}}$~=~2.1~GeV with the CEE Fast Simulation}\thanks{This work was supported in part by the National Key R\&D Program of China (No.~2024YFA1610700) and the National Natural Science Foundation of China (Grant No.~12375141).}

\author{Shuaike Lv}
\affiliation{State Key Laboratory of Particle Detection and Electronics, University of Science and Technology of China, Hefei 230026, China}
\affiliation{Department of Modern Physics, University of Science and Technology of China, Hefei 230026, China}

\author{Fan Si}
\affiliation{Physikalisches Institut, Ruprecht-Karls-Universit\"at Heidelberg, 69120 Heidelberg, Germany}

\author{Xionghong He}
\affiliation{Institute of Modern Physics, Chinese Academy of Sciences, Lanzhou 730000, China}

\author{Xu Sun}
\affiliation{Institute of Modern Physics, Chinese Academy of Sciences, Lanzhou 730000, China}

\author{Yige Huang}
\affiliation{Institute of Modern Physics, Chinese Academy of Sciences, Lanzhou 730000, China}

\author{Wanbing He}
\affiliation{Key Laboratory of Nuclear Physics and Ion-beam Application~(MOE),  Institute of Modern Physics, Fudan University, Shanghai 200433, China}

\author{Yifei Zhang}
\email{ephy@ustc.edu.cn}
\affiliation{State Key Laboratory of Particle Detection and Electronics, University of Science and Technology of China, Hefei 230026, China}
\affiliation{Department of Modern Physics, University of Science and Technology of China, Hefei 230026, China}

\author{Ming Shao}
\email{swing@ustc.edu.cn}
\affiliation{State Key Laboratory of Particle Detection and Electronics, University of Science and Technology of China, Hefei 230026, China}
\affiliation{Department of Modern Physics, University of Science and Technology of China, Hefei 230026, China}

\author{Ziyang Li}
\email{lzy9404@ustc.edu.cn}
\affiliation{State Key Laboratory of Particle Detection and Electronics, University of Science and Technology of China, Hefei 230026, China}
\affiliation{Department of Modern Physics, University of Science and Technology of China, Hefei 230026, China}

\begin{abstract}
The Cooling Storage Ring External-target Experiment (CEE) at the Heavy Ion Research Facility in Lanzhou (HIRFL) offers a unique opportunity to map the QCD phase diagram in the high-baryon-density region. In particular, higher-order cumulants ($C_r$) of event-by-event proton number distributions have been proposed as sensitive probes of the nuclear-matter phase structure. In this work, $^{\text{238}}$U+$^{\text{238}}$U collisions at $\sqrt{s_{\mathrm{NN}}}=$~2.1~GeV are generated using the JAM2 model and then processed through the CEE fast simulation with detector responses. Experimental analysis procedures, including centrality determination and particle identification, are also implemented. Proton-number cumulants up to fourth order and their ratios are calculated and compared with the true values. A track-by-track efficiency correction method is tested by introducing reconstruction efficiency within the TPC acceptance. An additional factor is also included to account for acceptance losses caused by the azimuthal gap between the two TPCs. Within the uncertainties of the current data sample, the corrected cumulants and cumulant ratios are consistent with their generator-level values. The event-sample dependence of the cumulant-ratio uncertainties is further investigated to estimate the statistics required for future measurements, revealing a substantial statistical cost associated with the acceptance compensation. These results provide a simulation-based assessment of the analysis feasibility and statistical requirements for proton-number fluctuation measurements at CEE.
\end{abstract}


\maketitle
\nolinenumbers

\section{Introduction}

One of the principal goals of nuclear physics is to explore the phase structure of strongly interacting matter, as summarized in the quantum chromodynamics (QCD) phase diagram characterized by temperature ($T$) and baryon chemical potential ($\mu_{\mathrm{B}}$)~\cite{Halasz:1998qr,Rajagopal:1999cp,Fukushima:2010bq}. The transition between the hadronic phase and the quark--gluon plasma (QGP) phase is of particular interest. Lattice QCD calculations indicate that this transition is a smooth crossover at vanishing or small $\mu_{\mathrm{B}}$~\cite{Aoki:2006we,Gupta:2011wh,HotQCD:2018pds}, whereas at larger $\mu_{\mathrm{B}}$, a first-order phase transition line terminating at a QCD critical endpoint (CEP) is predicted by effective models~\cite{Scavenius:2000qd,Hatta:2002sj,Antoniou:2002xq,Stephanov:2004wx}. Heavy-ion collisions provide an experimental approach to access different regions of the QCD phase diagram, by varying the center-of-mass energy per nucleon pair, $\sqrt{s_{\mathrm{NN}}}$. Collisions at high energies primarily probe matter at high temperature and low $\mu_{\mathrm{B}}$, while lower-energy collisions create matter with a larger net-baryon density corresponding to higher $\mu_{\mathrm{B}}$~\cite{Andronic:2017pug,Bzdak:2019pkr}.

Event-by-event fluctuations of conserved charges ($q$), including net-baryon number ($B$), net-electric-charge ($Q$) and net-strangeness ($S$), are considered sensitive probes of the QCD phase structure~\cite{Asakawa:2000wh,Hatta:2003wn}. In thermal equilibrium, cumulants $C_{r}$ of conserved-charge distributions and the corresponding QCD susceptibilities $\chi_{r}$ are directly related by $C_{n,q}=VT^{3}\chi_{n,q}$~\cite{Jeon:2003gk}. Ratios of cumulants cancel the trivial volume dependence and can therefore be compared with susceptibility ratios obtained from theoretical calculations and model estimates. Near the CEP, the growth of the correlation length ($\xi$) can strongly affect higher-order susceptibilities and non-Gaussian fluctuations~\cite{Stephanov:2008qz,Asakawa:2009aj}. Consequently, non-monotonic $\sqrt{s_{\mathrm{NN}}}$ dependence of higher-order cumulant ratios has been proposed as a characteristic experimental signature of the QCD CEP~\cite{Stephanov:2008qz,Stephanov:2011pb}.

Experimentally, due to the difficulty of event-by-event $B$ measurement, the net-proton-number ($\Delta N_{\mathrm{p}}=N_{\mathrm{p}}-N_{\mathrm{\bar{p}}}$) fluctuations are commonly employed as a proxy for the net-baryon number~\cite{Hatta:2003wn}. The STAR experiment has systematically measured net-proton-number fluctuations up to the fourth order in Au+Au collisions over a broad collision-energy range of $\sqrt{s_{\mathrm{NN}}}=7.7$--200~GeV within the Beam Energy Scan (BES) program at the Relativistic Heavy Ion Collider (RHIC)~\cite{STAR:2020tga,STAR:2021iop,STAR:2025zdq}. For 0--5\% central collisions, the energy dependence of the difference between the measured net-proton $\left.C_{4}\middle/C_{2}\right.$ and several reference baselines without CEP effects exhibits a minimum around $\sqrt{s_{\mathrm{NN}}}=$~19.6~GeV, where the deviations from the baselines have significances of approximately 2--5$\sigma$, depending on the reference. Another STAR measurement~\cite{STAR:2021fge,STAR:2022etb} presents the proton-number fluctuations up to the sixth order in Au+Au collisions at $\sqrt{s_{\mathrm{NN}}}=$~3~GeV, where the antiproton production is negligible. The $\left.C_{4}\middle/C_{2}\right.$ result is well reproduced by the UrQMD hadronic transport model~\cite{Bass:1998ca,Bleicher:1999xi} calculation, indicating that hadronic interactions dominate at such low energy. Proton-number fluctuations up to fourth order are measured at even-lower $\sqrt{s_{\mathrm{NN}}}=$~2.4~GeV by the HADES Collaboration~\cite{HADES:2020wpc}. It emphasizes the importance of studying the effect of volume fluctuations and light-nucleus formation at low energies in proton number fluctuation study. These measurements extend proton-number-fluctuation studies into the few-GeV energy regime and establish an important experimental basis for further fluctuation studies at low collision energies.

At the Heavy Ion Research Facility in Lanzhou (HIRFL), the Cooling Storage Ring (CSR)~\cite{Xia:2002xpu} delivers ion beams with kinetic energies ranging from several hundred MeV/u to about 1~GeV/u to the CSR External-target Experiment (CEE)~\cite{Lu:2016htm,Guo:2024zij}. CEE is a large-acceptance fixed-target magnetic spectrometer designed to investigate strongly interacting matter at high baryon densities. The proposed physics program includes $^{\text{238}}$U+$^{\text{238}}$U collisions at a beam kinetic energy of 500~MeV/u, corresponding to $\sqrt{s_{\mathrm{NN}}}=$~2.1~GeV and $\mu_{\mathrm{B}}\sim$~830~MeV~\cite{Cleymans:2005xv}. Measurements at CEE would therefore extend studies of the QCD phase diagram to higher net-baryon densities and enable systematic investigations of baryon-number conservation, event-by-event volume fluctuations, nuclear mean-field dynamics, and light-nucleus formation.

This work presents a feasibility study for the measurement of proton-number fluctuations in fixed-target $^{\text{238}}$U+$^{\text{238}}$U collisions at $\sqrt{s_{\mathrm{NN}}}=$~2.1~GeV. Events are generated with the second version of the Jet AA Microscopic Transport Model (JAM2)~\cite{Nara:jam2,Nara:1999dz,Nara:2021fuu} and subsequently processed through the CEE fast-simulation framework to model the detector acceptance and response. The analysis includes key experimental procedures, such as centrality determination and particle identification. The TPC reconstruction efficiency is estimated, and tracks are then randomly removed according to the corresponding inefficiency. The effects of finite reconstruction efficiency, incomplete azimuthal acceptance, and the bin-dependent PID selection on the proton-number cumulants are quantified, and the corresponding corrections are validated against their reference samples. In addition, the statistical uncertainties of the cumulant ratios are studied as functions of the total minimum-bias event sample to estimate the statistics required for a future CEE measurement.

\section{Event Generation and Fast Simulation}

\subsection{JAM2 Event Generation}

JAM2~\cite{Nara:jam2} is a Monte Carlo event generator based on a microscopic hadronic transport framework. In the JAM collision term, hadrons and resonances are propagated in space and time and undergo successive binary interactions and decays. Low-energy inelastic hadron--hadron interactions are described through resonance excitation and decay, whereas string excitation and fragmentation are employed above the resonance-dominated region~\cite{Nara:1999dz}. In addition to the cascade mode, JAM2 provides relativistic quantum molecular dynamics (RQMD) implementations in which baryons propagate under density- and momentum-dependent hadronic mean fields~\cite{Nara:2021fuu}. The RQMD mean-field mode is employed in the present study, rather than the pure cascade mode, so that the nuclear-medium effects during the transport evolution are explicitly included. JAM2 version 2.7273 is used throughout this work.

A total of $2\times10^{6}$ $^{\text{238}}$U+$^{\text{238}}$U collision events are generated at a projectile kinetic energy of 500~MeV/u, corresponding to $\sqrt{s_{\mathrm{NN}}}=2.1$~GeV. For this fixed-target configuration, the nucleon--nucleon center-of-mass frame has a rapidity of $y_{\mathrm{CM}}^{\mathrm{lab}}\simeq0.496$ in the laboratory frame. The particle rapidity in the center-of-mass frame is therefore given by $y_{\mathrm{CM}}=y-y_{\mathrm{CM}}^{\mathrm{lab}}$. The impact parameter ($b$) is sampled within 0--16~fm. The transport evolution is performed with a time-step size of $0.2$~fm/$c$ for 200 steps, corresponding to a total evolution time of 40~fm/$c$. Nuclear-cluster formation is included at the end of the transport evolution using a phase-space coalescence prescription with relative-distance and relative-momentum parameters $R_{0}=3.0$~fm and $P_{0}=0.3$~GeV/$c$, respectively. Nucleons assigned to the resulting clusters are removed from the final free-particle sample. The identities and four-momenta of the resulting final-state particles in each event are subsequently passed to the CEE fast-simulation chain. The same final-state sample before detector transport is retained as the generator-level reference for the comparisons below.

\subsection{CEE Detector and Fast-Simulation Framework}

\begin{figure}[t]
  \centering
  \subfigure[]{\includegraphics[width=\linewidth]{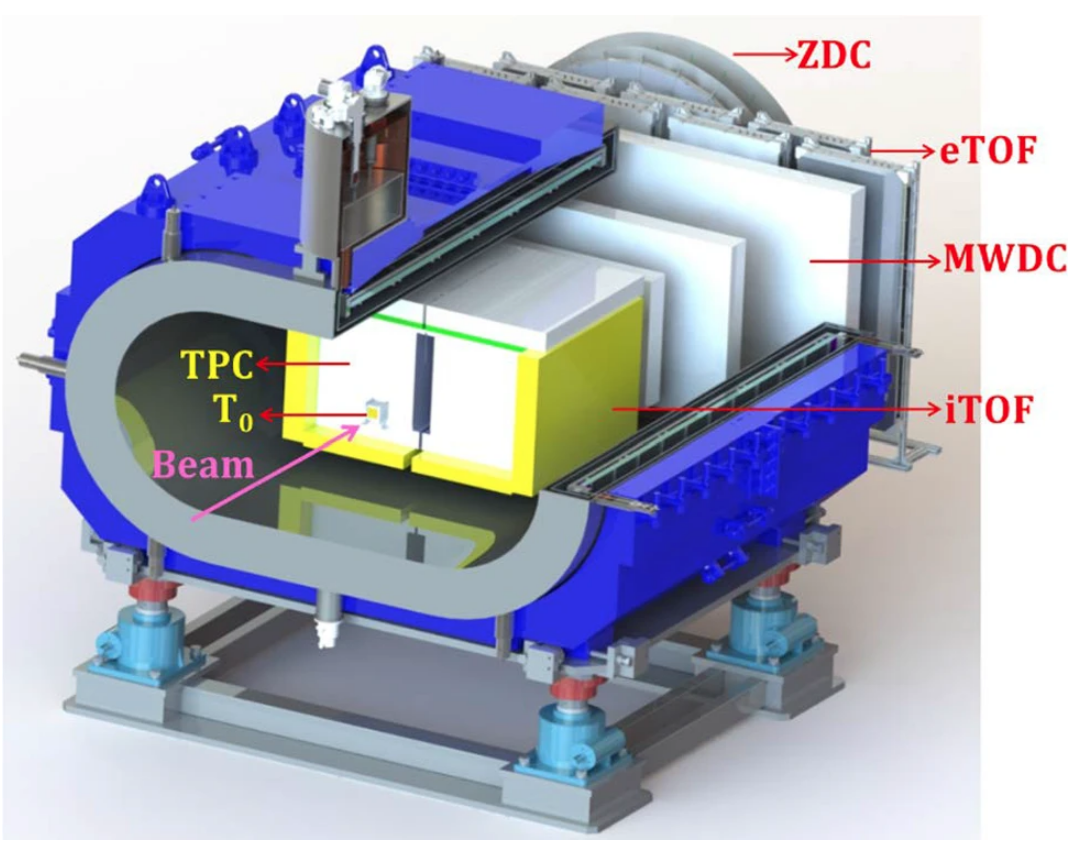}\label{fig:detector}}
  \par\smallskip
  \subfigure[]{\includegraphics[width=\linewidth]{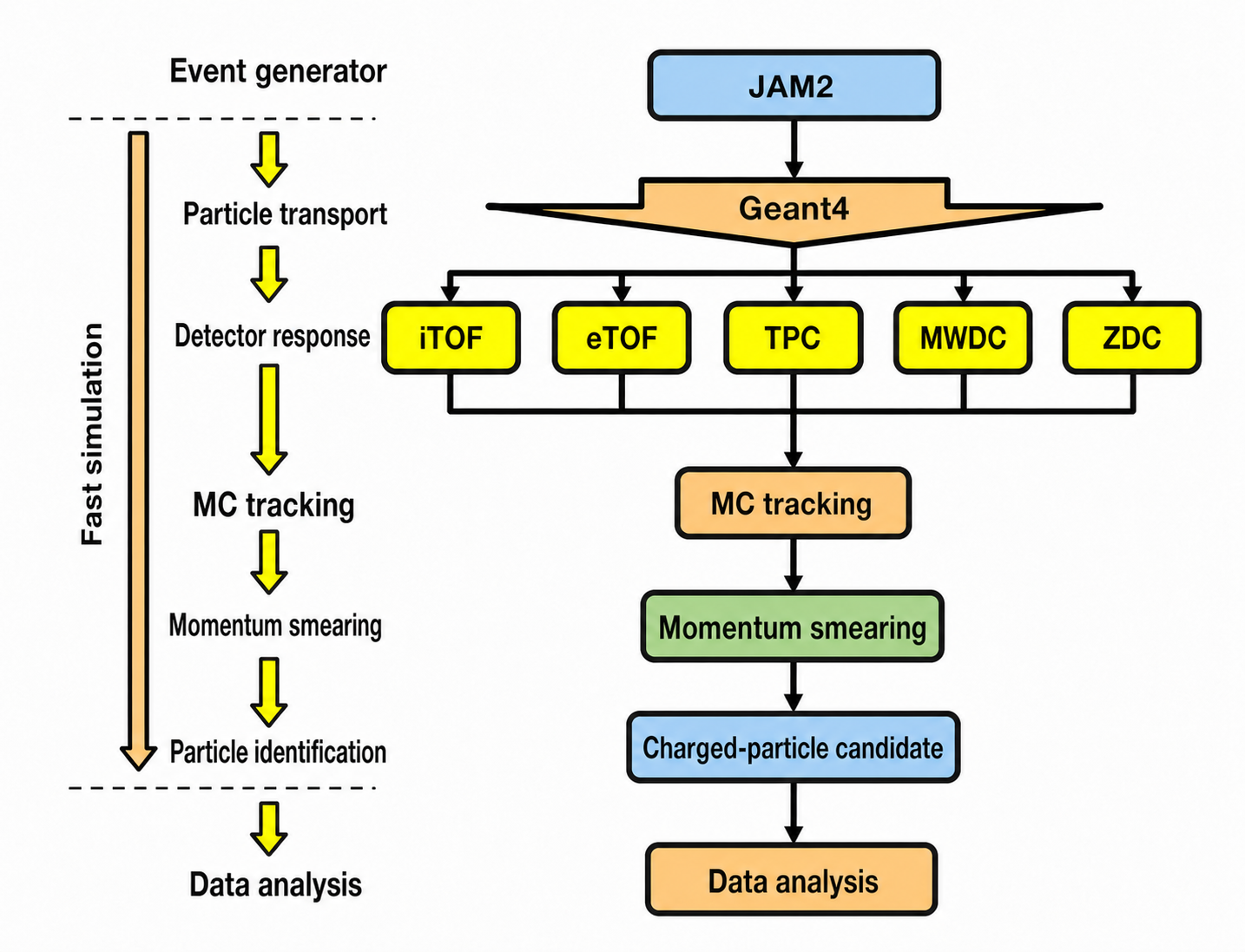}\label{fig:fastsim}}
  \caption{(a) Schematic layout of the CEE detector system~\cite{Lu:2016htm,Guo:2024zij}. The central TPCs provide the charged-particle tracks used for centrality estimation and proton-candidate selection. (b) Workflow of the CEE fast simulation~\cite{Guo:2024zij}. Events generated by the JAM2 model~\cite{Nara:jam2} undergo particle transport, detector-response modeling, MC tracking, momentum smearing, charged-particle candidate construction, and data analysis.}
  \label{fig:cee}
\end{figure}

The CEE detector system, shown schematically in \Cref{fig:detector}, is built around a large-gap superconducting dipole magnet that provides a nominal field of approximately 0.5~T~\cite{Lu:2016htm,Guo:2024zij}. Two TPCs~\cite{Huang:2018dus,Yang:2024sor}, surrounded by the inner time-of-flight (iTOF) system~\cite{Wang:2022uwc,Zhou:2022jsz}, cover the mid-rapidity region. At forward rapidity, three multi-wire drift chambers (MWDCs)~\cite{He:2024cew} are followed by the external time-of-flight (eTOF) wall~\cite{Wang:2020ehx}, while the zero-degree calorimeter (ZDC)~\cite{Zhu:2021soc} is located farther downstream. The start-time detector (T0)~\cite{Hu:2023niz} and a silicon-pixel beam monitor (BM)~\cite{Wang:2022evq} are installed along the beam line upstream of the target.

The beam direction defines the positive $z$-axis, and the positive $y$-axis points vertically upward, parallel to the nominal magnetic field. The two TPCs are located symmetrically about the beam line, leaving a central gap for beam passage. They provide measurements of charged-particle trajectory and specific ionization. Particle momenta are reconstructed from track curvature in the magnetic field, and the energy loss ($\mathrm{d}E/\mathrm{d}x$) response is used for particle identification. In the present analysis, TPC track candidates are used for both centrality estimation and proton-candidate selection.

The CEE fast-simulation framework~\cite{Guo:2024zij} is shown in \Cref{fig:fastsim}, where the detector response is evaluated. Final-state particles from JAM2 are propagated through the detector geometry and magnetic field with GEANT4 toolkit~\cite{GEANT4:2002zbu}, and their intersections with the active detector volumes are recorded. The subsequent detector-readout and reconstruction stages are simplified: full electronics digitization and data-like pattern recognition are not performed. Instead, simulated trajectories are processed with Monte Carlo truth association, while the relevant detector resolutions and reconstruction efficiencies are introduced through response parameterizations. Only the TPC branch of the framework is used in the present fluctuation analysis. The TPC geometry determines active-volume and pad-row crossings, while the parameterized response provides the reconstructed momentum, $\mathrm{d}E/\mathrm{d}x$. The resulting TPC candidates provide the charged-track multiplicity used for collision centrality determination and the proton-candidate identification used for the fluctuation analysis.

\section{Analysis Method}
\label{sec:AnaMeth}

\subsection{TPC Geometry and Reconstruction Efficiency}
\label{sec:tpc}

\begin{figure}[t]
  \centering
  \includegraphics[width=\linewidth]{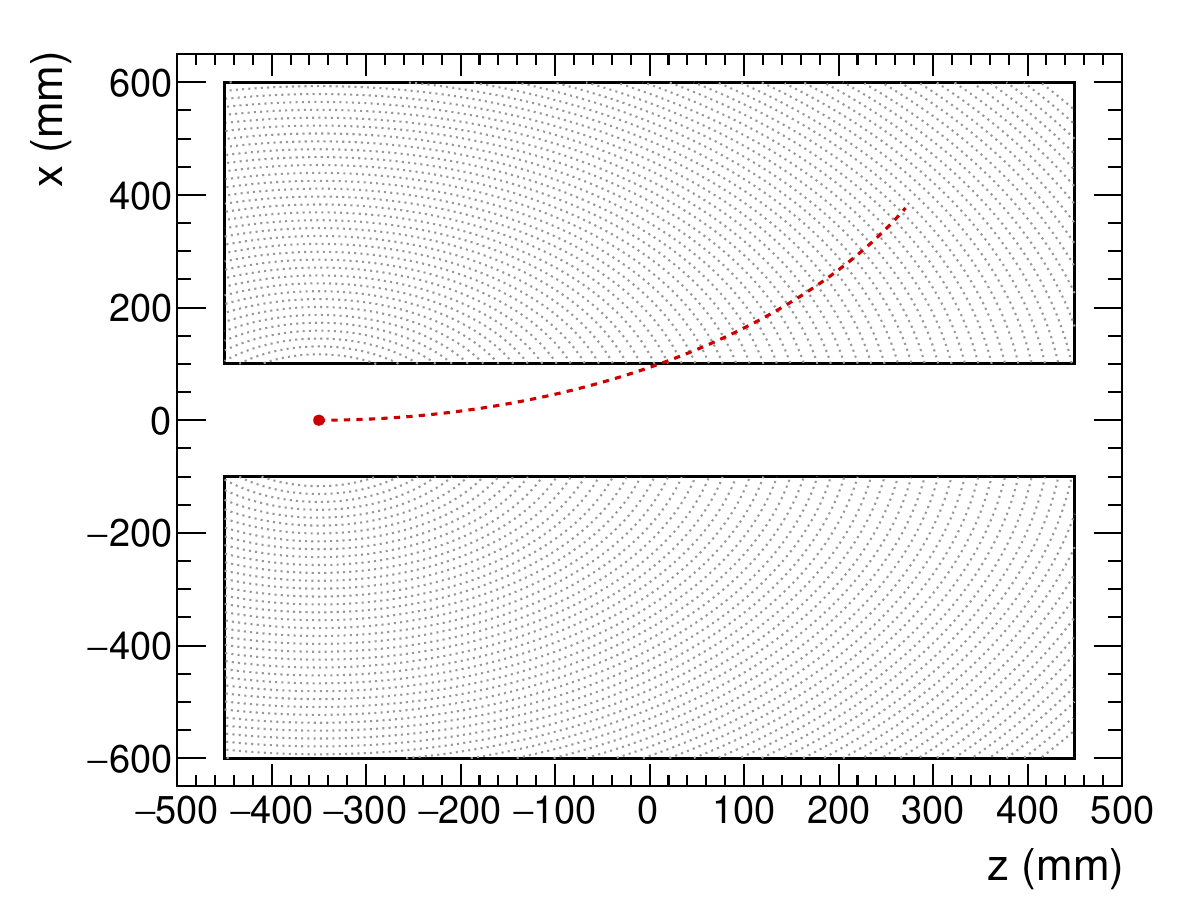}
  \caption{Schematic $x$-$z$ projection of a charged-particle trajectory crossing the two TPC active volumes. The red point marks the target, the dashed red curve represents the trajectory, and the internal lines indicate the pad-row boundaries used to determine the number of valid crossings.}
  \label{fig:XzRow}
\end{figure}

Since the magnetic field points along the $y$-axis, the projections of charged-particle trajectories bend in the $x$-$z$ plane, with curvature governed by the momentum component perpendicular to the field, $\sqrt{p_x^2+p_z^2}$, for a given charge and magnetic-field strength. In the $x$-$z$ projection illustrated in \Cref{fig:XzRow}, the active regions cover roughly $-450<z<450$~mm and $100<|x|<600$~mm. The beam passage and target opening consequently produce a nonuniform geometrical acceptance in azimuth. Each active volume comprises 62 pad rows, with the radial positions of their centers, measured from the target position, ranging approximately from 117 to 971~mm, as shown in \Cref{fig:XzRow}.

\begin{figure}[t]
  \centering
  \includegraphics[width=\linewidth]{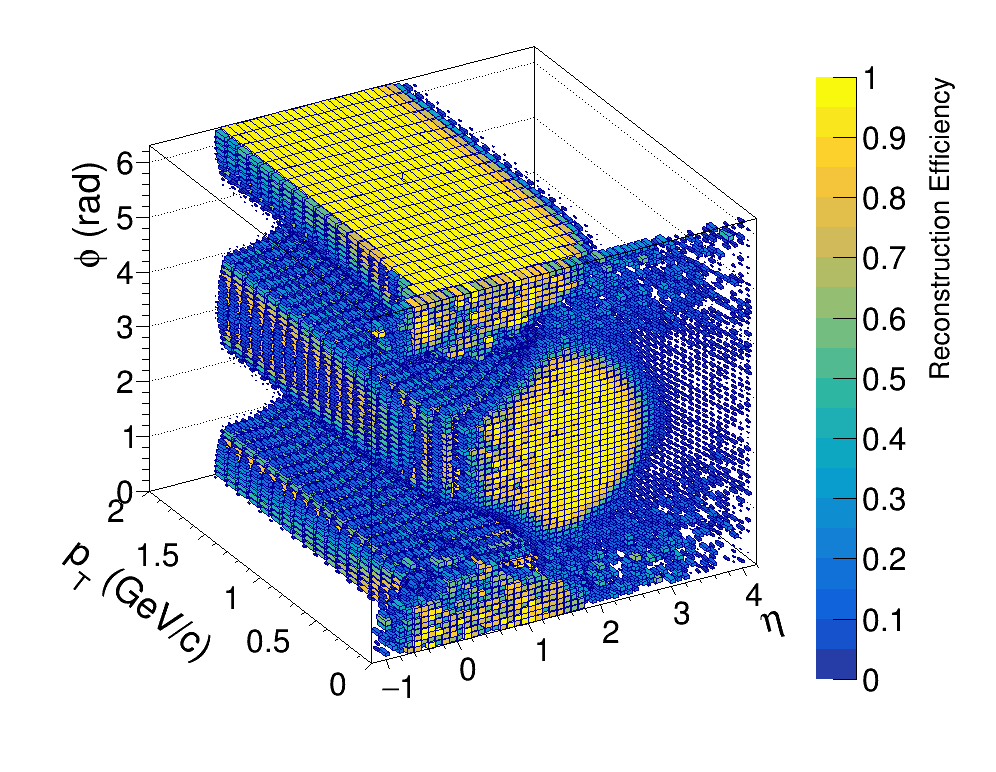}
  \caption{TPC reconstruction efficiency ($\varepsilon_{\mathrm{rec}}$) of protons in the three-dimensional phase space ($\eta$, $p_{\mathrm{T}}$, $\phi$). The color represents the average efficiency in each bin.}
  \label{fig:tpc_tracking_efficiency}
\end{figure}

\begin{figure}[t]
  \centering
  \subfigure[]{\includegraphics[width=\linewidth]{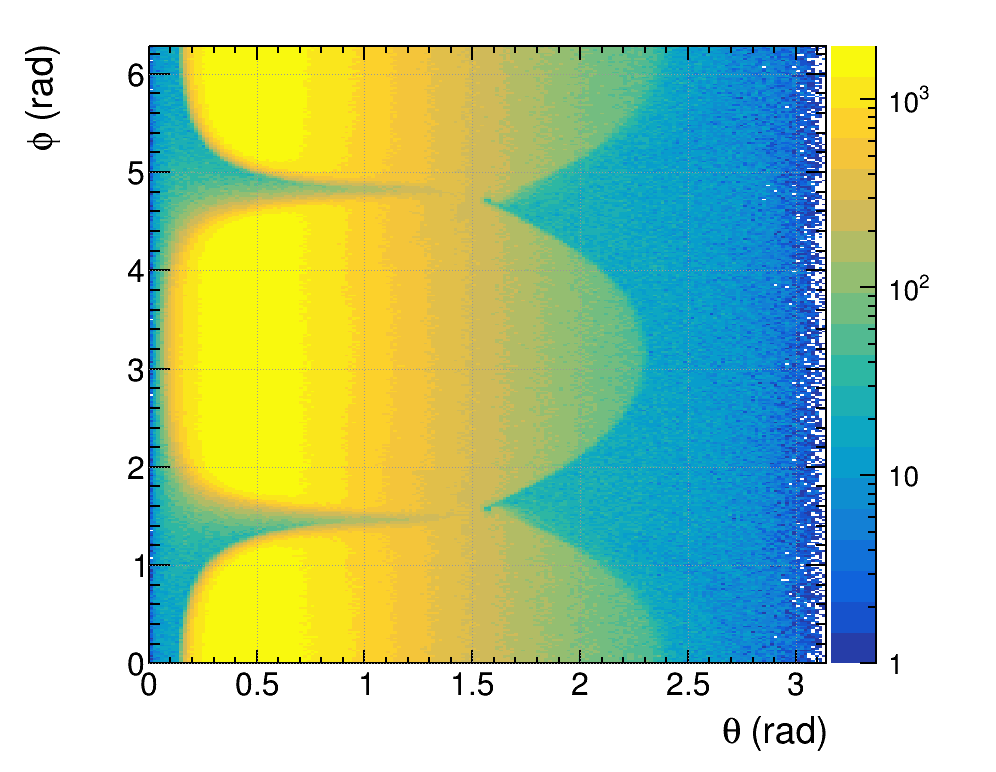}\label{fig:theta-phi}}
  \par\smallskip
  \subfigure[]{\includegraphics[width=\linewidth]{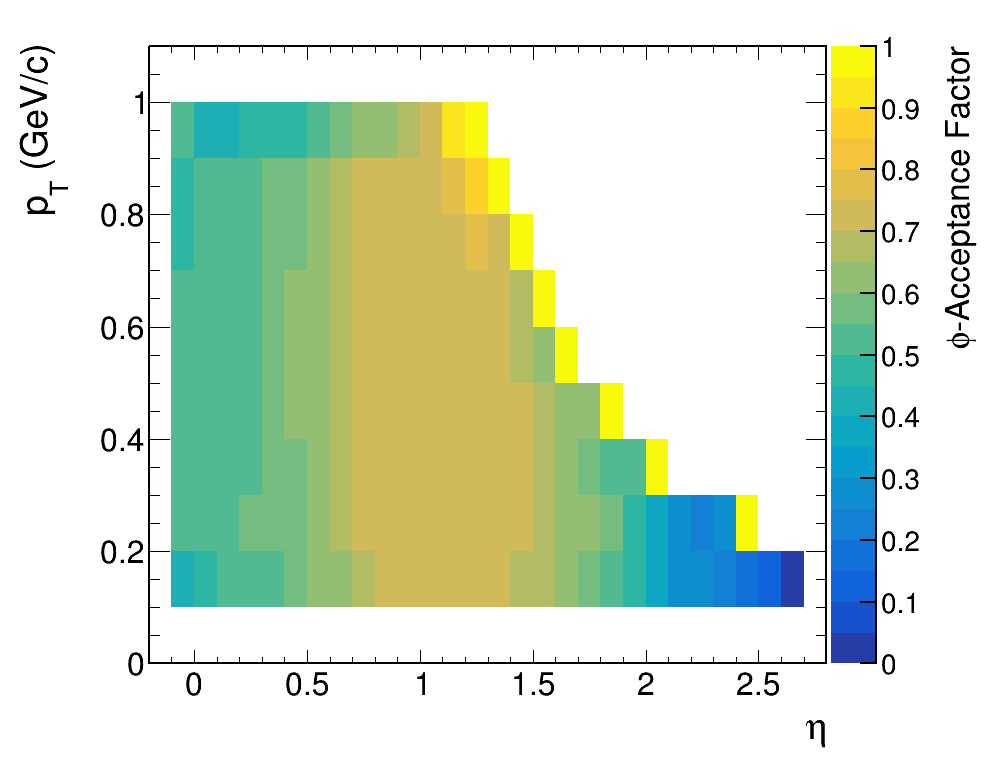}\label{fig:acceptance_factor}}
  \caption{(a) Distribution of TPC reconstructed protons ($N_{\mathrm{rec}}^{\mathrm{p}}$) in the $\theta$-$\phi$ plane. The absence of azimuthal coverage between the two TPCs contribute to the depleted bands. (b) $\phi$-acceptance factor ($\varepsilon_{\mathrm{acc}})$ in each ($\eta$, $p_{\mathrm{T}}$) bin. The color represents the fraction of generated protons within the $\phi$ bins of $\varepsilon_{\mathrm{rec}}>0.15$.}
\end{figure}

For a charged-particle trajectory entering a TPC active volume, the geometrical intersections determine the entry row, exit row, and the total number of crossed pad rows ($N_{\mathrm{cross}}$). Each intersected pad row defines a potential measurement position but contributes a space point to the track fit only when a valid hit is recorded and reconstructed. The total number of valid hits is denoted by $N_{\mathrm{hit}}$ ($\leq N_{\mathrm{cross}}$). The corresponding set of space points is used to fit the trajectory, determine its curvature, and reconstruct the particle momentum. The trajectory and $N_{\mathrm{cross}}$ of each charged particle can be determined by its four-momentum and electric charge and the magnetic field. Each crossed pad row is assumed to yield a valid hit independently with probability $p_{\mathrm{hit}}=0.6$, and the random $N_{\mathrm{hit}}$ of each track follows the binomial probability distribution $\mathrm{Binomial}(N_{\mathrm{cross}},p_{\mathrm{hit}})$. A track is considered reconstructible if at least $N_{\mathrm{hit}}^{\mathrm{cut}}=10$ valid hits are recorded. The corresponding conditional reconstruction efficiency is given by
\begin{equation}
\varepsilon_{\mathrm{rec}}=\sum_{k=N_{\mathrm{hit}}^{\mathrm{cut}}}^{N_{\mathrm{cross}}}
\binom{N_{\mathrm{cross}}}{k}p_{\mathrm{hit}}^{k}(1-p_{\mathrm{hit}})^{N_{\mathrm{cross}}-k},
\end{equation}
which represents the probability that a geometrically accepted trajectory satisfies the hit requirement. In general, we have
\begin{equation}
\varepsilon_{\mathrm{rec}}(\eta,p_{\mathrm{T}},\phi)
=
\frac{
N_{\mathrm{rec}}(\eta,p_{\mathrm{T}},\phi)
}{
N_{\mathrm{acc}}(\eta,p_{\mathrm{T}},\phi)
}.
\end{equation}

where $N_{\mathrm{acc}}$ denotes the number of particle candidates within the TPC geometrical acceptance that satisfy the common preselection criteria, before stochastic reconstruction losses are applied, and $N_{\mathrm{rec}}$ denotes the number of these candidates that are subsequently reconstructed. 

The loss of tracks in the TPC is therefore attributed to two distinct detector effects: limited geometrical acceptance and reconstruction inefficiency. \Cref{fig:tpc_tracking_efficiency} shows the TPC reconstruction efficiency of protons as a function of $\eta$, $p_{\mathrm{T}}$ and $\phi$. The three-dimensional dependence results from the TPC boundaries and the different curved trajectories in the magnetic field. Since low values of $\varepsilon_{\mathrm{rec}}$, resulting from the edge effect of the TPC geometry, would produce large uncertainties of high-order cumulants, phase-space bins with $\varepsilon_{\mathrm{rec}}<0.15$ are not included in the proton fluctuations.

The second source of track loss arises from the limited geometrical acceptance of the TPC. In the present detector configuration, this effect is dominated by the incomplete azimuthal coverage of the two TPCs, as shown in \Cref{fig:theta-phi}. To characterize the corresponding acceptance loss in the kinematic phase space used for the fluctuation analysis, the geometrical-acceptance efficiency is evaluated in the two-dimensional $(\eta,p_{\mathrm{T}})$ plane after integrating over $\phi$. \Cref{fig:acceptance_factor} shows the resulting proton acceptance factor as a function of $\eta$ and $p_{\mathrm{T}}$. For each $(\eta,p_{\mathrm{T}})$ bin, this factor quantifies the fraction of particles that fall within the azimuthal coverage of the TPC system. The geometrical-acceptance factor for a given particle species is defined by
\begin{equation}
\varepsilon_{\mathrm{acc}}(\eta,p_{\mathrm{T}})=
\frac{\int_{\varepsilon_{\mathrm{rec}}>0.15}N_{\mathrm{acc}}(\eta,p_{\mathrm{T}},\phi)\mathrm{d}\phi}{\int_{0}^{2\pi}N_{\mathrm{gen}}(\eta,p_{\mathrm{T}},\phi)\mathrm{d}\phi},
\end{equation}
so $1-\varepsilon_{\mathrm{acc}}$ characterizes the acceptance loss in azimuth. The distribution peaks in the central region of the acceptance and decreases toward forward $\eta$ and low $p_{\mathrm{T}}$, demonstrating that the azimuthal loss cannot be represented by a constant factor.

\subsection{Proton Identification and Phase-Space Selection}
\label{sec:proton_selection}

\begin{figure}[t]
  \centering
  \includegraphics[width=\linewidth]{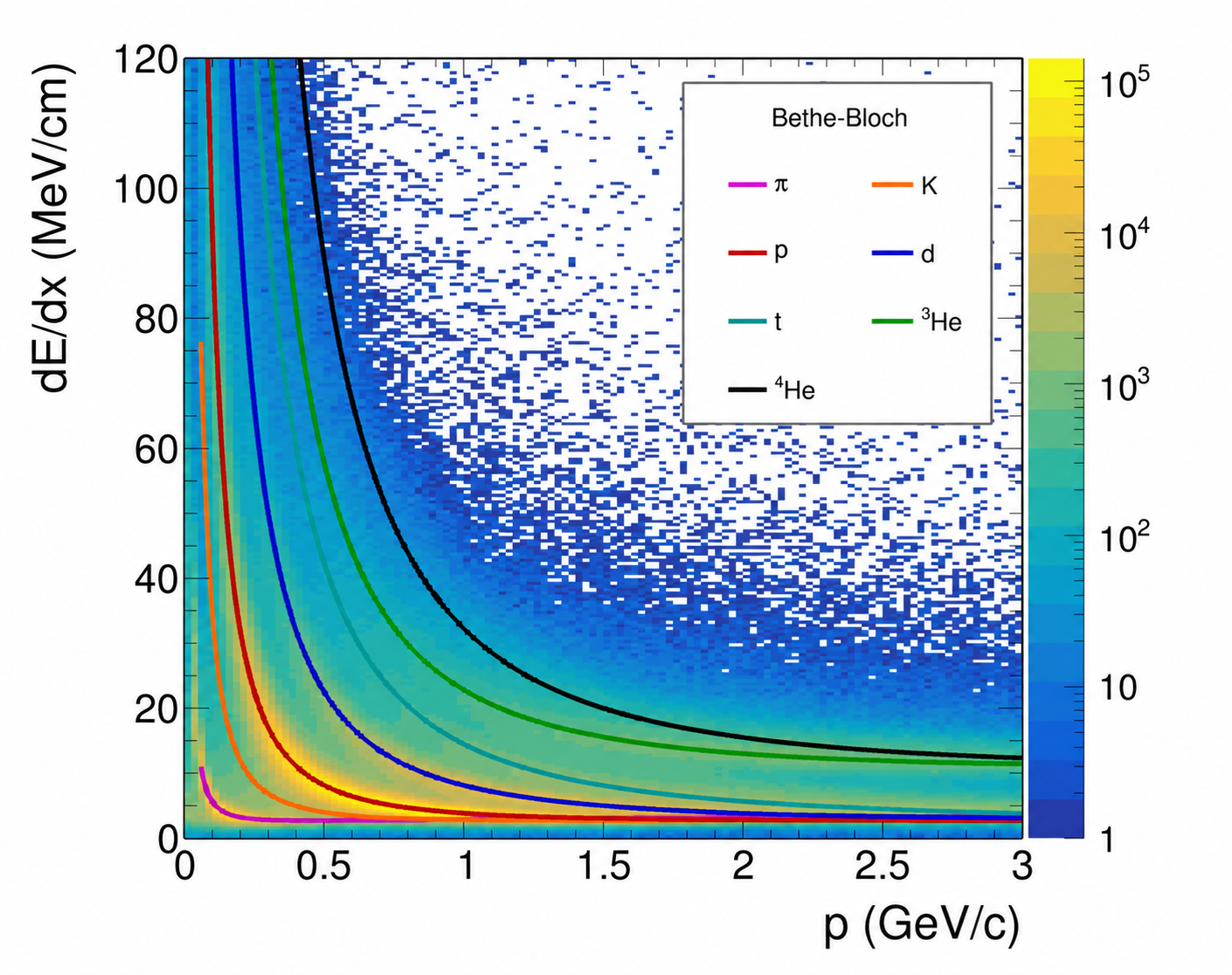}
  \caption{Distribution of the energy loss ($\mathrm{d}E/\mathrm{d}x$) in the TPC and the reconstructed total momentum ($p$) for the positively-charged tracks. The colored curves show the Bethe--Bloch expectations~\cite{Bethe:1930ku,Bloch:1933zza} for various particle species.}
  \label{fig:dedx}
\end{figure}

\begin{figure*}[t]
  \centering
  \subfigure[]{\includegraphics[width=0.48\textwidth]{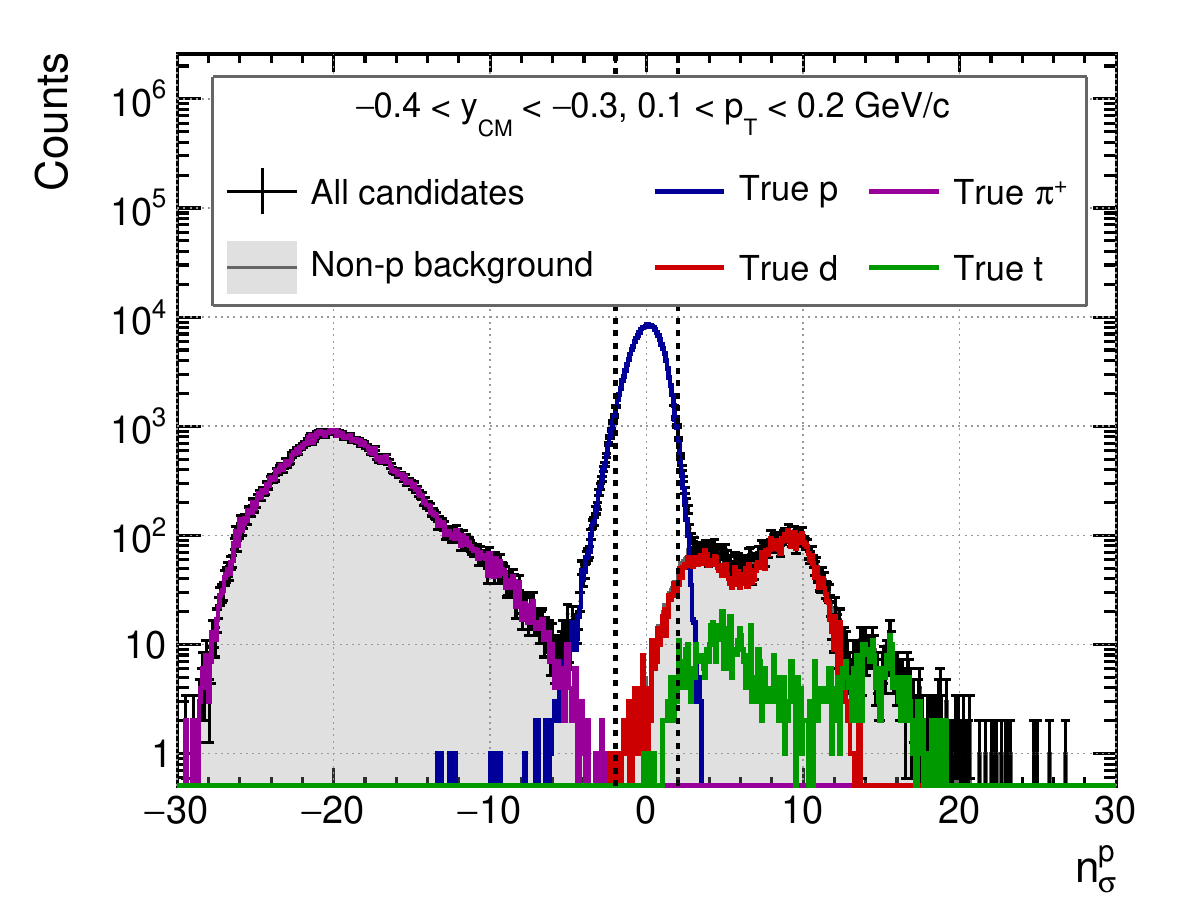}\label{fig:nsigma_low}}
  \subfigure[]{\includegraphics[width=0.48\textwidth]{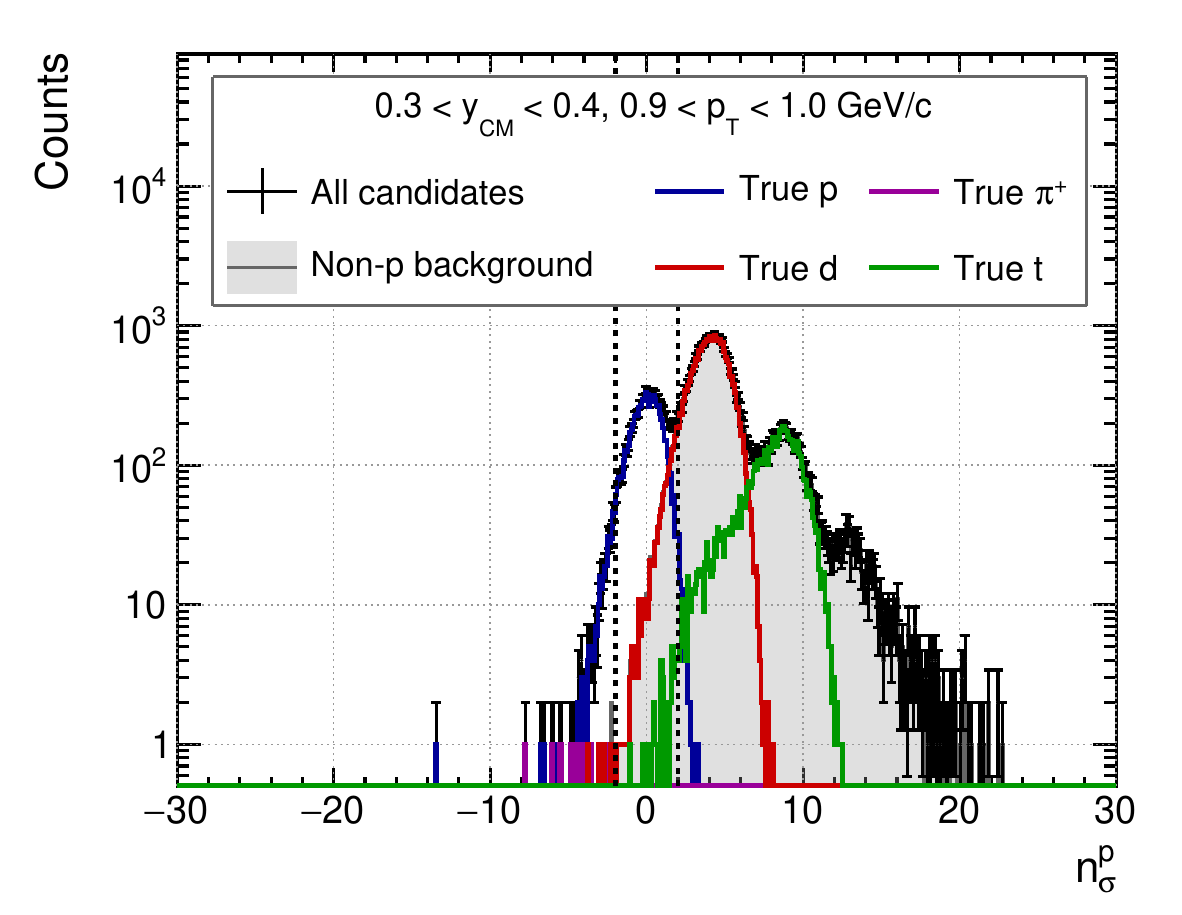}\label{fig:nsigma_high}}
  \par\medskip
  \subfigure[]{\includegraphics[width=0.48\textwidth]{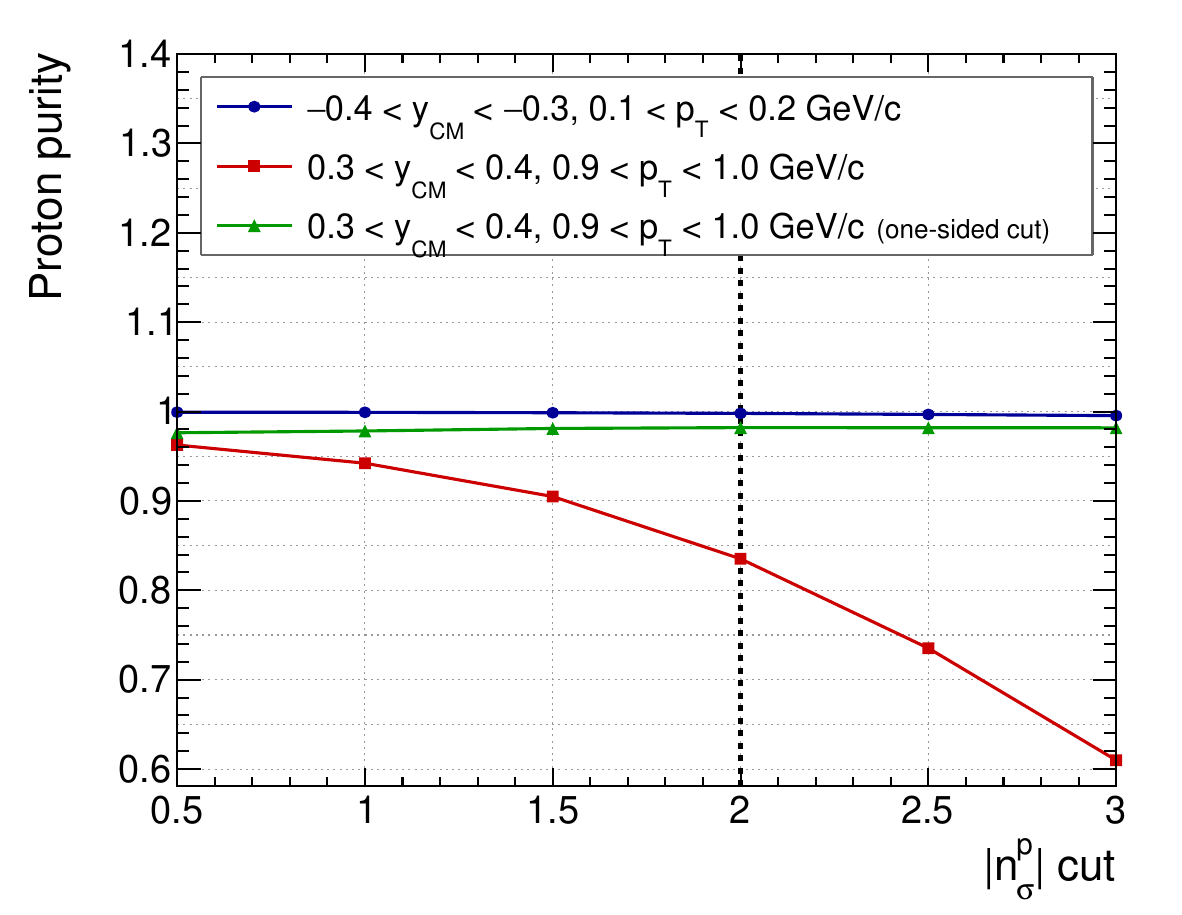}\label{fig:purity_typical}}
  \subfigure[]{\includegraphics[width=0.48\textwidth]{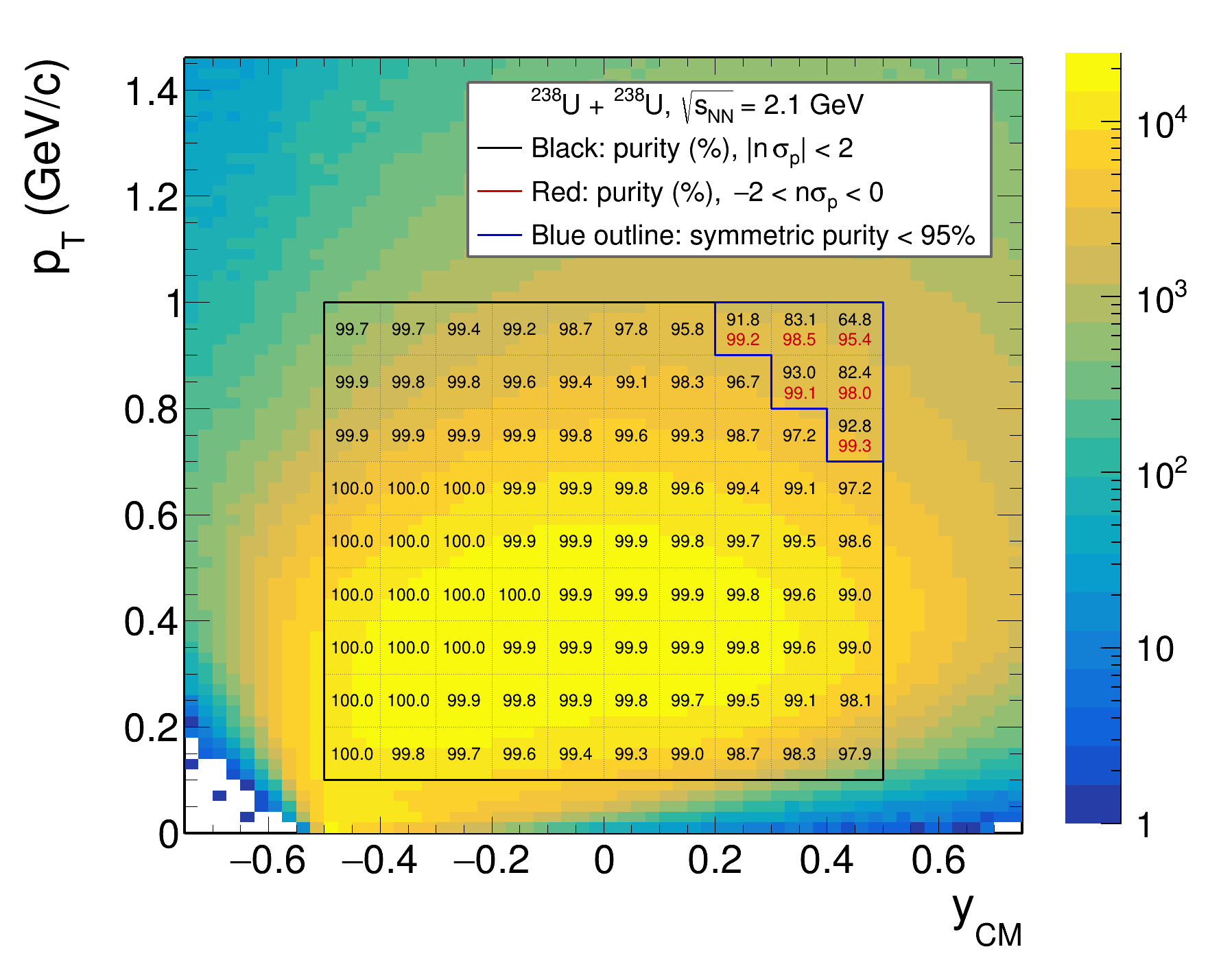}\label{fig:pid_purity_map}}
  \caption{PID performance for proton selection. (a) and (b) show the $n_{\sigma}^{\mathrm{p}}$ distributions within $-0.4<y_{\mathrm{CM}}<-0.3$ and $0.1<p_{\mathrm{T}}<0.2$~GeV/$c$, and within $0.3<y_{\mathrm{CM}}<0.4$ and $0.9<p_{\mathrm{T}}<1.0$~GeV/$c$, respectively. Black distributions include all positively charged tracks, blue curves represent true protons, and other colored curves and gray bands show the non-proton contributions. (c) Proton purities as functions of the $n_{\sigma}^{\mathrm{p}}$ cut ($c$), where $\left|n_{\sigma}^{\mathrm{p}}\right|<c$ (symmetric) or $-c<n_{\sigma}^{\mathrm{p}}<0$ (one-sided). (d) Distribution of ($y_{\mathrm{CM}}$, $p_{\mathrm{T}}$) of proton candidates (colored band) and proton purity (numbers in percent) in analysis bins, where $y_{\mathrm{CM}}$ is calculated under the proton-mass hypothesis. Black numbers within the black block give the purity obtained with the default symmetric selection $\left|n_{\sigma}^{\mathrm{p}}\right|<2$. The blue block marks bins in which this purity is below 95\%, and red numbers give the purity with the one-sided selection $-2<n_{\sigma}^{\mathrm{p}}<0$.}
  \label{fig:pid_performance}
\end{figure*}

\Cref{fig:dedx} shows the correlation between the track $\mathrm{d}E/\mathrm{d}x$ in the TPC and reconstructed total momentum ($p_{\mathrm{}}$). The curves denote the Bethe--Bloch expectations~\cite{Bethe:1930ku,Bloch:1933zza}, $\left(\mathrm{d}E/\mathrm{d}x\right)_{\mathrm{th}}^{\mathrm{X}}$, for different particle species $\mathrm{X}$. Their momentum dependence produces distinct PID bands in the low-momentum region.

The PID cut for protons is based on the quantity $n_{\sigma}^{\mathrm{p}}$ defined by
\begin{equation}
n_{\sigma}^{\mathrm{p}}=\frac{1}{\sigma_{\mathrm{d}E/\mathrm{d}x}}\ln\frac{\mathrm{d}E/\mathrm{d}x}{\left(\mathrm{d}E/\mathrm{d}x\right)_{\mathrm{th}}^{\mathrm{p}}}
\end{equation}
where $\sigma_{\mathrm{d}E/\mathrm{d}x}$ is the relative logarithmic $\mathrm{d}E/\mathrm{d}x$ resolution. For pure protons, this normalization derives a distribution centered near zero with a width close to unity. The $n_{\sigma}^{\mathrm{p}}$ distribution is checked within different two-dimensional phase space bins with widths of $\Delta y_{\mathrm{CM}}=0.1$, where $y_{\mathrm{CM}}$ is obtained using the proton mass for all tracks, and $\Delta p_{\mathrm{T}}=0.1$~GeV/$c$. \Cref{fig:nsigma_low,fig:nsigma_high} show examples of a low-momentum bin ($-0.4<y_{\mathrm{CM}}<-0.3$ and $0.1<p_{\mathrm{T}}<0.2~\mathrm{GeV}/c$) and a high-momentum bin ($0.3<y_{\mathrm{CM}}<0.4$ and $0.9<p_{\mathrm{T}}<1.0~\mathrm{GeV}/c$), respectively. The black and blue distributions represent positively-charged tracks and protons, respectively, and other colored curves show the contributions from contaminating particles with positive charges. The particle types are known from MC truth. The two vertical dashed lines in each panel characterize the window of $\left|n_{\sigma}^{\mathrm{p}}\right|<2$. At low momentum, the narrow proton peak at zero is clearly separated from other particles, and the contamination is negligible near $n_{\sigma}^{\mathrm{p}}=0$. The deuteron and triton components exhibit pronounced split structures. This splitting originates from the stronger curvature of low-momentum tracks together with the finite TPC acceptance. Tracks with different trajectories can have substantially different effective path lengths when they are truncated by the detector boundaries, leading to differences in the reconstructed momentum and the sampled $\mathrm{d}E/\mathrm{d}x$ response. At high momentum, the deuteron and triton contributions instead form broader single peaks on the positive side of the proton peak. As the momentum increases, the helix radius increases approximately as $r\propto p_{xz}/B$, reducing the track curvature. Consequently, the effective path-length distributions of different track populations overlap more strongly, and the separated $n_{\sigma}^{\mathrm{p}}$ components observed at low $p_{\mathrm{T}}$ merge into a single broader peak for each isotope. The two dominant structures on the positive side are therefore associated with deuterons and tritons, respectively.

The proton purity in each phase space bin can be given by
\begin{equation}
p(y_{\mathrm{CM}},p_{\mathrm{T}},W)=\frac{\int_{W}N_{\mathrm{rec}}^{\mathrm{p}}(y_{\mathrm{CM}},p_{\mathrm{T}},n_{\sigma}^{\mathrm{p}})\mathrm{d}n_{\sigma}^{\mathrm{p}}}{\int_{W}N_{\mathrm{rec}}^{\mathrm{h}^{+}}(y_{\mathrm{CM}},p_{\mathrm{T}},n_{\sigma}^{\mathrm{p}})\mathrm{d}n_{\sigma}^{\mathrm{p}}},
\end{equation}
where $N_{\mathrm{rec}}^{\mathrm{p}}$ and $N_{\mathrm{rec}}^{\mathrm{h}^{+}}$ represent the reconstructed numbers of MC true protons and positively-charged tracks, respectively, and $W$ is the $n_{\sigma}^{\mathrm{p}}$ cut window. \Cref{fig:purity_typical} shows the proton purity as a function of $\left|n_{\sigma}^{\mathrm{p}}\right|$ cut , where $\left|n_{\sigma}^{\mathrm{p}}\right|<c$, in the two representative bins: blue for low momentum in \Cref{fig:nsigma_low} and red for high momentum in \Cref{fig:nsigma_high}. As the selection window is widened, the proton purity gradually decreases because more reconstructed background tracks are included. In the high-momentum bin, the purity falls below 95\% at $c=1$, mainly due to contamination from reconstructed light nuclei with $Z=+1$. However, since the light-nucleus background is predominantly located on the positive side of the $n_{\sigma}^{\mathrm{p}}$ distribution shown in \Cref{fig:nsigma_high}, it is effective to strongly suppress the contamination by applying an asymmetric, one-sided cut of $-c<n_{\sigma}^{\mathrm{p}}<0$, and the proton purity is improved from the red line to the green line.

\Cref{fig:pid_purity_map} shows the kinematic map of proton candidates as the colored band. The black rectangle defines the analysis range: $\left|y_{\mathrm{CM}}\right|<0.5$ and $0.1<p_{\mathrm{T}}<1.0~\mathrm{GeV}/c$. In each cell, a black number is printed as the proton purity of $\left|n_{\sigma}^{\mathrm{p}}\right|<2$, which provides at least 95\% purity over most of the phase space. The purity decreases at higher momentum, and there are six cells surrounded by blue lines with the purity below 95\%. The one-sided $n_{\sigma}^{\mathrm{p}}$ cut of $-2<n_{\sigma}^{\mathrm{p}}<0$ is applied and forms the red purity numbers higher than 95\%.

An additional component of efficiency, the PID cut efficiency as a function of $y_{\mathrm{CM}}$ and $p_{\mathrm{T}}$, should also be included, which is expressed by
\begin{equation}
\varepsilon_{\mathrm{pid}}(y_{\mathrm{CM}},p_{\mathrm{T}})=\frac{\int_{W(y_{\mathrm{CM}},p_{\mathrm{T}})}N_{\mathrm{rec}}^{\mathrm{p}}(y_{\mathrm{CM}},p_{\mathrm{T}},n_{\sigma}^{\mathrm{p}})\mathrm{d}n_{\sigma}^{\mathrm{p}}}{N_{\mathrm{rec}}^{\mathrm{p}}(y_{\mathrm{CM}},p_{\mathrm{T}})},
\end{equation}
where $W(y_{\mathrm{CM}},p_{\mathrm{T}})$ represents the dynamic $n_{\sigma}^{\mathrm{p}}$ cut window exactly used in the corresponding kinematic cell and could be either symmetric or one-sided. It characterizes the fraction of the protons retained by the $n_{\sigma}^{\mathrm{p}}$ cut.

\subsection{Centrality Determination}

Collision centrality characterizes the geometrical overlap of the colliding nuclei and could be defined by $b$ or the number of participating nucleons ($N_{\mathrm{part}}$) generated in the model simulation, neither of which is accessible in experiments. In this analysis, the centrality estimator is defined by the number of reconstructed charged tracks, applicable in future experimental data. Only charged tracks surviving the detector-response simulation and reconstructed in the TPC are considered. A track-quality requirement of $N_{\mathrm{hit}}>10$ is imposed. Thus, the detector acceptance and reconstruction inefficiency are naturally included.

\begin{figure}[t]
  \centering
  \includegraphics[width=\linewidth]{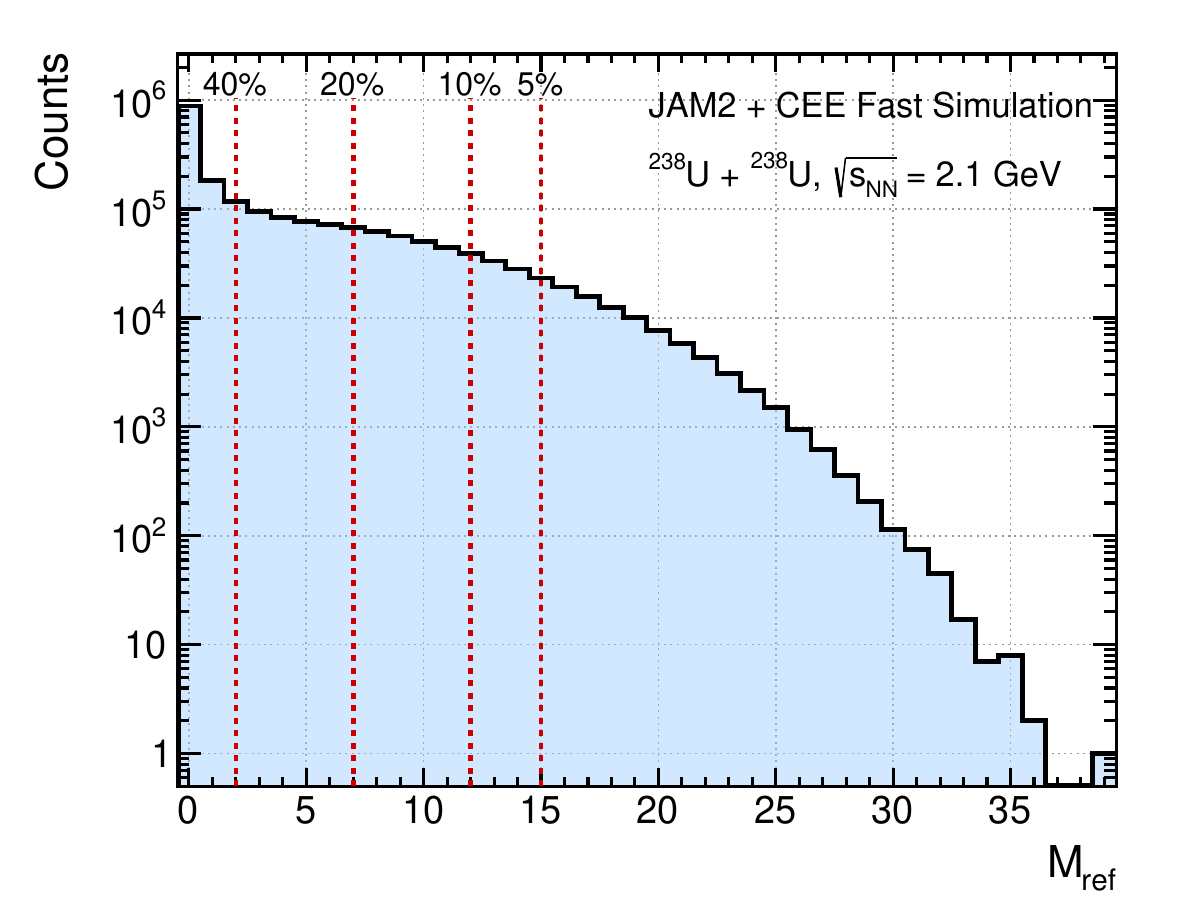}
  \caption{Distribution of the reference multiplicity of charged particles excluding protons ($M_{\mathrm{ref}}$). The detector-level $M_{\mathrm{ref}}$ includes the response of finite acceptance , reconstruction inefficiency and PID. Dashed lines indicate the 0\%--5\%, 5\%--10\%, 10\%--20\%, and 20\%--40\% centrality boundaries.}
  \label{fig:centrality}
\end{figure}

To reduce auto-correlations~\cite{Luo:2013bmi} between the centrality estimator and the proton number $N_{\mathrm{p}}$, tracks identified as proton candidates are excluded from the centrality estimator. The resulting reference charged-particle multiplicity is therefore defined as
\begin{equation}
M_{\mathrm{ref}}=N_{\mathrm{ch}}\left(N_{\mathrm{hit}}>10,\ \left|n_{\sigma}^{\mathrm{p}}\right|>2\right),
\end{equation}
where $N_{\mathrm{ch}}$ denotes the number of charged tracks. The $M_{\mathrm{ref}}$ distribution is shown in \Cref{fig:centrality}. It is integrated from high to low multiplicities, and centrality classes are defined by the percentiles. \Cref{tab:centrality} summarizes the $M_{\mathrm{ref}}$ divisions for 0\%--5\%, 5\%--10\%, 10\%--20\% and 20\%--40\% centrality classes. The boundaries are indicated by the dashed lines in \Cref{fig:centrality}.

\begin{table}[t]
    \caption{Summary of reference charged-particle multiplicity
    ($M_{\mathrm{ref}}$) ranges for various centrality classes.}
    \centering
    \begin{threeparttable}
        \begin{tabular}{|c|c|}
            \hline
            \makebox[0.20\columnwidth][c]{Centrality} & \makebox[0.75\columnwidth][c]{$M_{\mathrm{ref}}$ Ranges} \\
            \hline
            \makebox[0.20\columnwidth][c]{0\%--5\%} & \makebox[0.75\columnwidth][c]{$M_{\mathrm{ref}}>15$} \\
            \hline
            5\%--10\% & $12<M_{\mathrm{ref}}\leq15$ \\
            \hline
            10\%--20\% & $7<M_{\mathrm{ref}}\leq12$ \\
            \hline
            20\%--40\% & $2<M_{\mathrm{ref}}\leq7$ \\
            \hline
        \end{tabular}
    \end{threeparttable}
    \label{tab:centrality}
\end{table}

\subsection{Proton-Number Cumulants}

The basic event observable is the number of proton candidates ($N_{\mathrm{p}}$) satisfying the selection criteria summarized in \Cref{tab:selection}. No explicit antiproton number is constructed because antiproton production is negligible at this beam energy. The observable throughout this work is therefore the proton-number cumulant. With $\delta N_{\mathrm{p}}=N_{\mathrm{p}}-\langle N_{\mathrm{p}}\rangle$, where $\langle\cdot\rangle$ derives the mean, the first four cumulants are expressed by
\begin{align}
C_1 &= \left\langle N_{\mathrm{p}}\right\rangle,\\
C_2 &= \left\langle(\delta N_{\mathrm{p}})^2\right\rangle,\\
C_3 &= \left\langle(\delta N_{\mathrm{p}})^3\right\rangle,\\
C_4 &= \left\langle(\delta N_{\mathrm{p}})^4\right\rangle-3\left\langle(\delta N_{\mathrm{p}})^2\right\rangle^2.
\end{align}
Here, $C_1$ is the mean proton number, $C_2$ is the variance, $C_3$ characterizes the asymmetry of the distribution, and $C_4$ is sensitive to non-Gaussian tails. The ratios $C_2/C_1$, $C_3/C_2$ and $C_4/C_2$ are also studied. They reduce the trivial system-size dependence but do not automatically remove detector effects, centrality resolution or finite-statistics effects.

\begin{table}[t]
    \caption{Proton selection criteria.}
    \centering
    \begin{threeparttable}
        \begin{tabular}{|c|c|}
            \hline
            \makebox[0.20\columnwidth][c]{Quality} & \makebox[0.75\columnwidth][c]{$N_{\mathrm{hit}}\geq10$} \\
            \hline
            \multirow{3}{*}{Kinematics} & $\left|y_{\mathrm{CM}}\right|<0.5$ \\
            & $0.1<p_{\mathrm{T}}<1.0~\mathrm{GeV}/c$ \\
            & $\varepsilon_{\mathrm{rec}}>0.15$ \\
            \hline
            \multirow{2}{*}{PID} & $\left|n_{\sigma}^{\mathrm{p}}\right|<2$, if $p\left(\left|n_{\sigma}^{\mathrm{p}}\right|<2\right)\geq95\%$ \\
            & $-2<n_{\sigma}^{\mathrm{p}}<0$, otherwise \\
            \hline
        \end{tabular}
    \end{threeparttable}
    \label{tab:selection}
\end{table}

\subsection{Efficiency and Acceptance Corrections}
\label{sec:efficiency_correction}

The TPC reconstruction efficiency and PID cut efficiency randomly retain a fraction of proton tracks within the kinematic range and consequently distort the observed event-by-event $N_{\mathrm{p}}$ distribution and its cumulants. We apply the track-by-track method based on the binomial response~\cite{Nonaka:2017kko,Luo:2018ofd,Si:2021mdj} to test the closure of this random loss. An acceptance correction is also tested separately with the $\phi$-acceptance factor considered as an efficiency for proton numbers within the geometric acceptance with $\phi$ gaps and correct their cumulants to those within a full azimuthal range. The idea is to use the tracks within the acceptance to represent those outside the acceptance.

For a constant binomial efficiency ($\varepsilon$), the probability distribution of the observed multiplicity ($n$) is derived by
\begin{equation}
P_{\mathrm{obs}}(n)=\sum_{N=n}^{\infty}P_{\mathrm{true}}(N)B_{N,\varepsilon}(n)
\end{equation}
where $P_{\mathrm{true}}(N)$ denotes the true-multiplicity probability distribution, and
\begin{equation}
B_{N,\varepsilon}(n)=\binom{N}{n}\varepsilon^n(1-\varepsilon)^{N-n}.
\end{equation}
It is proved that the $r$th-order factorial cumulant satisfies~\cite{Nonaka:2017kko}
\begin{equation}
\langle n^r\rangle_{\mathrm{fc}}=\varepsilon^r\langle N^r\rangle_{\mathrm{fc}}.
\end{equation}
When the efficiency varies across tracks, the track-by-track method~\cite{Luo:2018ofd} labels the single-track efficiency $\varepsilon_{j}$ for the track indexed by $j$ and defines the event-wise quantity
\begin{equation}
q_{r,s}=\sum_j\frac{a_j^r}{\varepsilon_j^s},
\end{equation}
where the sum runs over proton candidates, and the charge $a_j=1$. Denoting cumulants over the event ensemble by $\langle\cdot\rangle_{\mathrm{c}}$, the first two corrected cumulants are
\begin{align}
C_1^{\mathrm{corr}}&=\langle q_{1,1}\rangle_{\mathrm{c}},\\
C_2^{\mathrm{corr}}&=\langle q_{1,1}^{2}\rangle_{\mathrm{c}}+\langle q_{2,1}\rangle_{\mathrm{c}}-\langle q_{2,2}\rangle_{\mathrm{c}}.
\end{align}
Higher-order cumulants are corrected through the corresponding relations in Ref.~\cite{Nonaka:2017kko}. Note that the method assumes a binomial response and an accurate single-track efficiency, and correlated track loss, occupancy effects or an imperfect efficiency model can therefore prevent complete closure.

For the efficiency correction, we use
\begin{equation}
\varepsilon_{j}=\varepsilon_{\mathrm{rec}}(\eta_j,p_{\mathrm{T},j},\phi_j)\varepsilon_{\mathrm{pid}}(y_{\mathrm{CM},j},p_{\mathrm{T},j}).
\end{equation}
The test is applied to correct for the random loss due to track reconstruction, including the proton identification. For efficiency--acceptance joint correction, we use
\begin{equation}
\varepsilon_{j}=\varepsilon_{\mathrm{rec}}(\eta_j,p_{\mathrm{T},j},\phi_j)\varepsilon_{\mathrm{pid}}(y_{\mathrm{CM},j},p_{\mathrm{T},j})\varepsilon_{\mathrm{acc}}(\eta_j,p_{\mathrm{T},j}).
\end{equation}
Because $\varepsilon_{\mathrm{acc}}$ is taken from an average over $\phi$ rather than defined locally in the uncovered phase space, the compensation assumes that the covered azimuth statistically represents the full azimuthal range within each $(\eta,p_{\mathrm{T}})$ bin. It is therefore a feasibility test dependent on azimuthal symmetry and the simulated response, not a model-independent extrapolation.

\subsection{Centrality-Bin-Width Correction}

Each broad centrality class contains events with several integer $M_{\mathrm{ref}}$ values, and the system volume varies inside. To reduce the dependence of the cumulants on this finite bin width, the Centrality-Bin-Width Correction (CBWC)~\cite{Luo:2013bmi,STAR:2021iop} is applied. In each finest centrality class, the single-valued $M_{\mathrm{ref}}$ bin indexed by $i$, the $r$th-order cumulant ($C_{r,i}$) is first calculated and then averaged over the broad class with the bin-by-bin number of events ($N_{\mathrm{evt},i}$) as the weight:
\begin{equation}
C_r^{\mathrm{CBWC}}=\frac{\sum_i N_{\mathrm{evt},i}C_{r,i}}{\sum_i N_{\mathrm{evt},i}}.
\end{equation}
The same treatment of weighted average from fine bins is also applied consistently to cumulant ratios. In this work, the CBWC is used to reduce the mixing introduced by finite centrality-bin widths; it is not assumed to remove all participant or volume fluctuations at low collision energy.

\subsection{Statistical Uncertainty Estimation}

Statistical uncertainties are estimated with the analytical error-propagation method~\cite{Stuart:2010ats,Luo:2011tp,Si:2021vog}. For moment estimators $\hat{m}_r$ and $\hat{m}_s$ evaluated from identical $N_{\mathrm{evt}}$ events, the covariance is given by
\begin{equation}
N_{\mathrm{evt}}\operatorname{Cov}(\hat{m}_r,\hat{m}_s)=m_{r+s}-m_rm_s.
\end{equation}
Using the relations between moments and cumulants, the variances of cumulant estimators up to the fourth order can be derived as
\begin{align}
N_{\mathrm{evt}}\operatorname{Var}(\hat{C}_1)&=C_2,\\
N_{\mathrm{evt}}\operatorname{Var}(\hat{C}_2)&=C_4+2C_2^2,\\
N_{\mathrm{evt}}\operatorname{Var}(\hat{C}_3)&=C_6+9C_4C_2+9C_3^2+6C_2^3,\\
\begin{split}
N_{\mathrm{evt}}\operatorname{Var}(\hat{C}_4)&=C_8+16C_6C_2+48C_5C_3+34C_4^2\\
&\quad+72C_4C_2^2+144C_3^2C_2+24C_2^4.
\end{split}
\end{align}
The fourth-order uncertainty contains information up to eighth order and is consequently much larger than those of $C_1$ and $C_2$ for finite samples.

For efficiency-corrected cumulant estimators, the $q_{r,s}$ variables and their products are treated as correlated multivariate random variables. The first two variances are
\begin{align}
N_{\mathrm{evt}}\operatorname{Var}(\hat{C}_1^{\mathrm{corr}})&=\langle q_{(1,1)}^2\rangle_{\mathrm{c}},\\
\begin{split}
N_{\mathrm{evt}}\operatorname{Var}(\hat{C}_2^{\mathrm{corr}})&=\langle q_{(1,1)}^{4}\rangle_{\mathrm{c}}+2\langle q_{(1,1)}^{2}\rangle_{\mathrm{c}}^{2}+2\langle q_{(1,1)}^{2} q_{(2,1)}\rangle_{\mathrm{c}}\\
&\quad-2\langle q_{(1,1)}^{2} q_{(2,2)}\rangle_{\mathrm{c}}+\langle q_{(2,1)}^{2}\rangle_{\mathrm{c}}\\
&\quad-2\langle q_{(2,1)} q_{(2,2)}\rangle_{\mathrm{c}}+\langle q_{(2,2)}^{2}\rangle_{\mathrm{c}},
\end{split}
\end{align}
and those for higher orders can be found in Ref.~\cite{Si:2021vog}.

The variance of the cumulant ratio is also obtained from the error propagation as
\begin{equation}
\operatorname{Var}(\hat{C}_r/\hat{C}_s)=\frac{\operatorname{Var}(\hat{C}_r)}{C_s^2}+\frac{C_r^2\operatorname{Var}(\hat{C}_s)}{C_s^4}-\frac{2C_r\operatorname{Cov}(\hat{C}_r,\hat{C}_s)}{C_s^3}
\end{equation}
for both uncorrected and corrected levels.

In the CBWC, events from fine $M_{\mathrm{ref}}$ bins are statistically independent, and statistical uncertainty are propagated through the weighted average by
\begin{equation}
\operatorname{Var}(\hat{C}_r^{\mathrm{CBWC}})=
\frac{\sum_i N_{\mathrm{evt},i}^2\operatorname{Var}(\hat{C}_{r,i})}{(\sum_i N_{\mathrm{evt},i})^2}.
\end{equation}

\section{Results and Discussion}

\subsection{Cumulants and Cumulant Ratios}

\begin{figure}[b]
  \centering
  \includegraphics[width=\hsize]{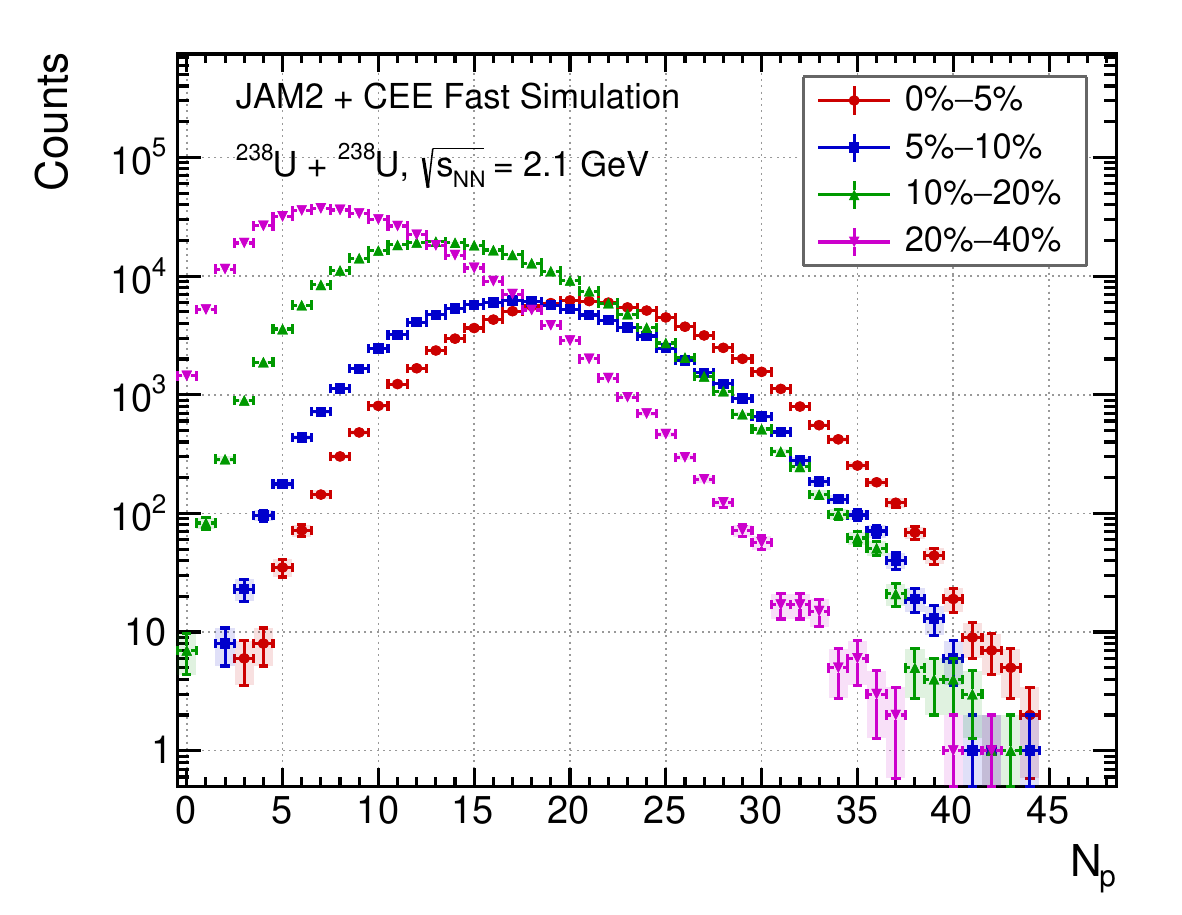}
  \caption{Proton number ($N_{\mathrm{p}}$) distributions in the four centrality classes (0\%--5\%, 5--10\%, 10--20\% and 20\%--40\%).}
  \label{fig:proton_num}
\end{figure}

\Cref{fig:proton_num} shows the event-by-event proton number ($N_{\mathrm{p}}$) distributions in the four centrality classes. Here, $N_{\mathrm{p}}$ denotes the number of reconstructed proton candidates after detector effects and the applied proton-selection criteria, rather than the true generator-level proton multiplicity.  From peripheral to central collisions, the distributions shift toward larger proton number and become broader, reflecting the increase in system size and mid-rapidity proton production. These distributions constitute the direct inputs to the cumulant calculation and illustrate the different proton number scales and fluctuation widths sampled in the four centrality intervals.

Based on these proton number distributions, we next examine the corresponding proton-number cumulants and quantify how they are modified by the detector response. The validation is carried out in two steps. First, the track-by-track efficiency correction is tested against the proton sample within the TPC geometrical acceptance. Second, the additional $\phi$-averaged acceptance correction is compared with the generator-level proton sample in the same kinematic window. This two-step comparison separates the effects of reconstruction inefficiency and incomplete azimuthal coverage and allows their respective impacts on the proton cumulants to be evaluated.

\begin{figure*}[!tbp]
  \centering
  \includegraphics[width=0.98\textwidth]{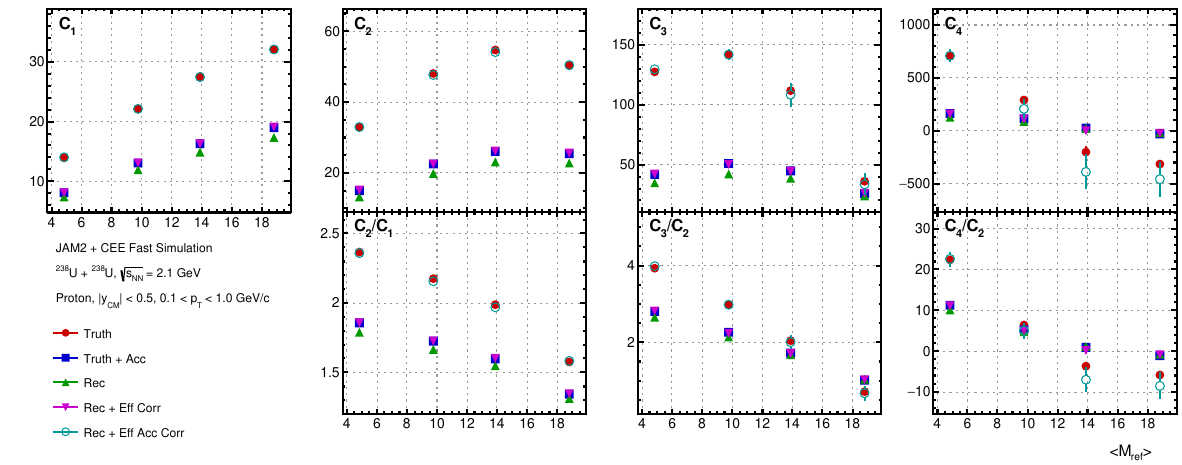}
  \caption{Centrality dependence of the proton-number cumulants and cumulant ratios as functions of $\langle M_{\mathrm{ref}}\rangle$. The upper row shows $C_1$, $C_2$, $C_3$, and $C_4$, while the lower row shows $C_2/C_1$, $C_3/C_2$, and $C_4/C_2$. All analysis levels use the same kinematic ($y_{\mathrm{CM}}$, $p_{\mathrm{T}}$) window, centrality definition and CBWC.}
  \label{fig:cumulants}
\end{figure*}

Five analysis levels are used to distinguish acceptance effects, reconstruction and PID losses and their corresponding corrections:

\begin{itemize}

\item \textit{Truth}: generator-level true proton-number cumulants calculated within the common ($y_{\mathrm{CM}}$, $p_{\mathrm{T}}$) window, before introducing any detector response.

\item \textit{Truth + Acc}: generator-level true proton-number cumulants calculated within the TPC acceptance satisfying the common ($y_{\mathrm{CM}}$, $p_{\mathrm{T}}$) window and the phase space cells with $\varepsilon_{\mathrm{rec}}>0.15$. This sample provides an acceptance-limited reference.

\item \textit{Rec}: uncorrected cumulants of proton candidates within the same kinematic range as \textit{Truth + Acc} after applying the TPC reconstruction response and the PID with the the $n_{\sigma}^{\mathrm{p}}$ cut window $W(y_{\mathrm{CM}},p_{\mathrm{T}})$. The associated proton losses are induced by the finite $\varepsilon_{\mathrm{rec}}$ and $\varepsilon_{\mathrm{pid}}$, respectively.

\item \textit{Rec + Eff Corr}: proton-number cumulants corrected for reconstruction and PID losses using $\varepsilon_{\mathrm{rec}}\varepsilon_{\mathrm{pid}}$. Comparison with \textit{Truth + Acc} tests the closure of the efficiency correction within the TPC acceptance.

\item \textit{Rec + Eff Acc Corr}: proton-number cumulants obtained by additionally including the $\phi$-acceptance factor $\varepsilon_{\mathrm{acc}}(\eta,p_{\mathrm{T}})$. The combined correction therefore uses $\varepsilon_{\mathrm{rec}}\varepsilon_{\mathrm{pid}}\varepsilon_{\mathrm{acc}}$ and is compared with \textit{Truth} to assess recovery of the full-azimuth reference within the common kinematic window.

\end{itemize}

All five levels use the same kinematic window of $\left|y_{\mathrm{CM}}\right|<0.5$ and $0.1<p_{\mathrm{T}}<1.0~\mathrm{GeV}/c$, centrality classes determined by $M_{\mathrm{ref}}$, and CBWC procedure. \Cref{fig:cumulants} summarizes the centrality dependence of cumulants up to the fourth order in upper panels and their ratios in lower panels. As the class becomes central, $C_1$ monotonically grows, and both $C_2$ and $C_3$ increase and then decrease. However, $C_4$ and three ratios always decrease from peripheral to central events. The comparison among analysis levels is discussed in the following.

The $C_1$ and $C_2$ results show the clearest hierarchy. The separation between \textit{Truth} (red circles) and \textit{Truth + Acc} (blue squares) is primarily caused by the incomplete azimuthal coverage and the detector-level candidate requirements. Random tracking loss due to the inefficiency of reconstruction and PID produces a further reduction of \textit{Rec} (green triangles). After the track-by-track correction, cumulants and cumulant ratios of \textit{Rec + Eff Corr} (magenta inverted triangles) return to the \textit{Truth + Acc} reference over the full centrality range. The additional acceptance correction brings the results of \textit{Rec + Eff Acc Corr} (teal open circles) close to \textit{Truth}. The hierarchy could be roughly described as: \textit{Rec} $<$ \textit{Rec + Eff Corr} $\sim$ \textit{Truth + Acc} $<$ \textit{Rec + Eff Acc Corr} $\sim$ \textit{Truth}. This simultaneous recovery of corrections is nontrivial for $C_2$, because the random detection distorts both the multiplicity and the sampling variance rather than merely rescaling the mean. Consistently, the corrected $C_2/C_1$ in the lower row follows the corresponding reference, while retaining its clear decrease toward larger $\langle M_{\mathrm{ref}}\rangle$. The track loss is not canceled by taking the cumulant ratio but corrected by the efficiency correction. The additional acceptance correction reproduces the generator-level centrality dependence within the present precision

The third-order results exhibit the same correction sequence but with greater sensitivity to the distribution tails. The efficiency-corrected $C_3$ and $C_3/C_2$ remain close to \textit{Truth + Acc}, showing that the correction also closes for third order under the imposed response. With the acceptance correction included, $C_3$ and $C_3/C_2$ follow \textit{Truth}, although the uncertainties become larger particularly in two most central classes. The hierarchy for the last point of $C_3/C_2$ becomes: \textit{Truth} $<$ \textit{Truth + Acc} $\sim$ \textit{Rec}, which is different from the second-order case.

For $C_4$ and $C_4/C_2$, the generator reference changes from positive values at small $\langle M_{\mathrm{ref}}\rangle$ to negative values at larger $\langle M_{\mathrm{ref}}\rangle$. The efficiency-corrected points follow those of \textit{Truth + Acc}, which are close to zero. The efficiency-and-acceptance-corrected result reproduces the sign change and the overall evolution of \textit{Truth}. The last two corrected points carry large error bars touching the true values.

Overall, the combined comparison shows good closure of the efficiency correction toward \textit{Truth + Acc} from the first to fourth orders. The additional $\phi$-averaged acceptance correction provides a good agreement with \textit{Truth}, especially for lower orders. The statistical uncertainties are visibly enlarged by the acceptance correction for higher orders in central events, which could be further studied using higher statistics.

\subsection{Statistical Precision}
\label{sec:statistical_precision}

\begin{figure}[!t]
  \centering
  \includegraphics[width=0.95\hsize]{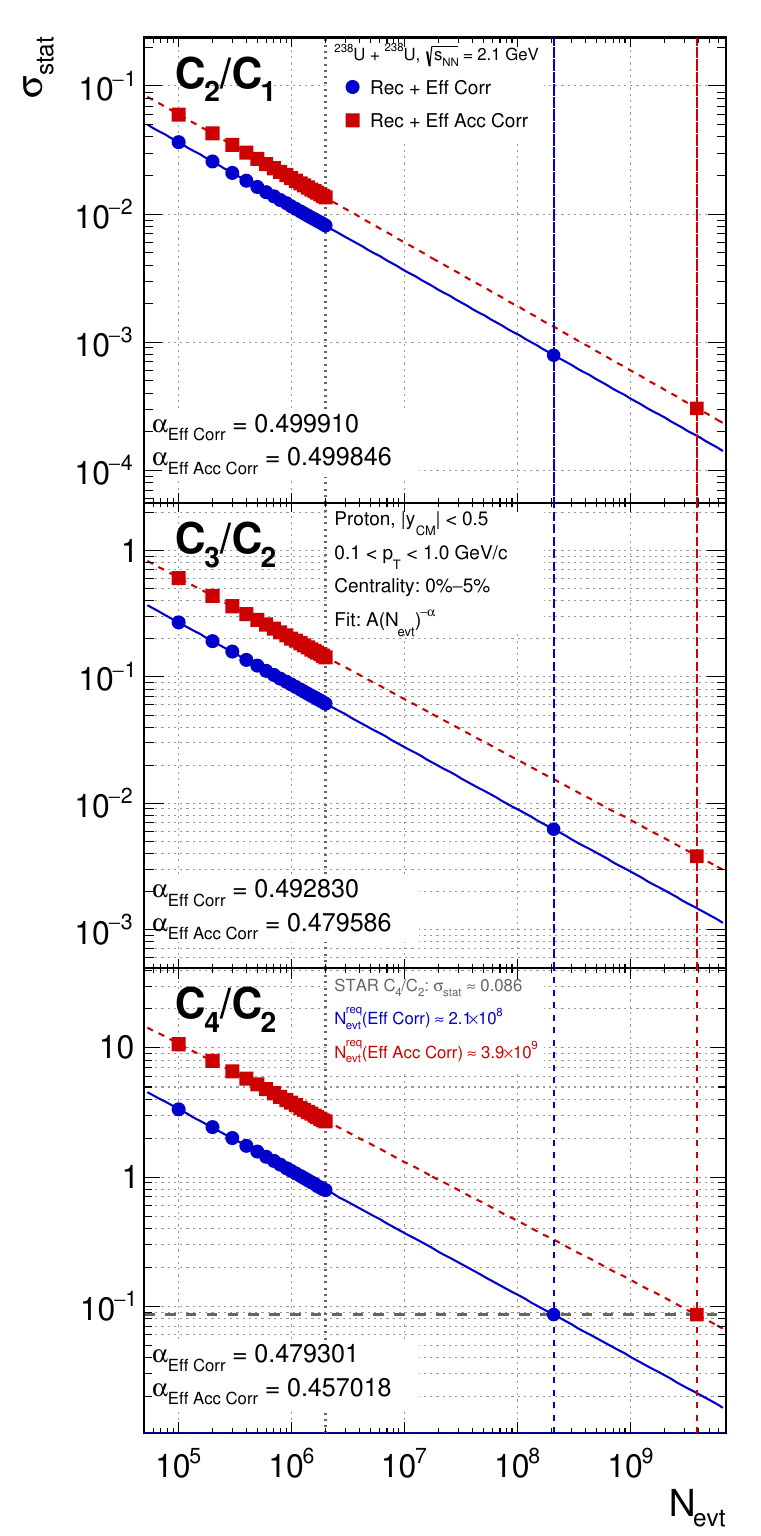}
  \caption{Statistical uncertainties of the proton-number cumulant ratios
  ($C_2/C_1$, $C_3/C_2$ and $C_4/C_2$) in 0\%--5\% central
  $^{238}$U+$^{238}$U collisions as functions of the total number of
  minimum-bias events. Blue circles show the \textit{Rec + Eff Corr} results,
  and red squares display the results after the \textit{Rec + Eff Acc Corr}.
  The solid lines are power-law fits for corresponding colors. In the
  $C_4/C_2$ panel, the horizontal dashed line marks a reference statistical
  uncertainty of 0.086, and the vertical dashed lines indicate the
  corresponding minimum-bias event requirements inferred from the fits.}
  \label{fig:statistical_precision}
\end{figure}

The enlarged uncertainties of higher-order cumulants motivate an estimate of the event sample required for a future CEE measurement. \Cref{fig:statistical_precision} shows the absolute statistical uncertainties of $C_2/C_1$, $C_3/C_2$ and $C_4/C_2$ in the 0\%--5\% centrality class as functions of the total number of minimum-bias events. The blue series uses the efficiency correction, whereas the red series is based on the combined efficiency--acceptance correction.

The statistical uncertainty of the cumulant ratio ($R$) is parametrized by
\begin{equation}
\sigma_{\mathrm{stat}}(R)=A\left(N_{\mathrm{evt}}^{\mathrm{}}\right)^{-\alpha},
\end{equation}
where $A$ and $\alpha$ are free parameters. For $C_2/C_1$, the fitted exponents are $\alpha=0.49991$ and $0.49985$ for the \textit{Rec + Eff Corr} and \textit{Rec + Eff Acc Corr} results, respectively, both consistent with the ideal $1/\sqrt{N_{\mathrm{evt}}}$ scaling. For $C_3/C_2$, the corresponding exponents are $0.49283$ and $0.47959$, while for $C_4/C_2$ they decrease further to $0.47930$ and $0.45702$. The fitted exponents for the higher-order cumulant ratios are slightly below the ideal value of 0.5 over the present event-statistics range. This deviation may reflect finite-sample effects, since higher-order cumulants are more sensitive to sparsely populated distribution tails and to fluctuations amplified by efficiency-correction weights. However, the deviations from 0.5 are modest, and their impact on the present statistical-precision estimates is expected to be limited. With sufficiently large event samples, the statistical uncertainties are generally expected to approach the asymptotic $1/\sqrt{N_{\mathrm{evt}}}$ behavior.

As a practical precision benchmark, the horizontal gray dashed line in the $C_4/C_2$ panel is set to $\sigma_{\mathrm{stat}}(C_4/C_2)\simeq0.086$, approximately corresponding to the statistical uncertainty achieved for the 0\%--5\% central net-proton measurement in the STAR Au+Au data at $\sqrt{s_{\mathrm{NN}}}=3$~GeV~\cite{STAR:2022etb}. This value is used only as a target statistical uncertainty rather than for a direct comparison of the physics results, because the experimental setup differs from CEE. Extrapolating the fitted power laws gives minimum-bias event requirements of approximately $2.1\times10^{8}$ for the reconstruction-efficiency-corrected result and $3.9\times10^{9}$ for the combined efficiency--acceptance correction. At $N_{\mathrm{evt}}\simeq2.1\times10^{8}$, the corresponding statistical uncertainties for the efficiency-only correction are $\sigma_{\mathrm{stat}}(C_2/C_1)\simeq8.0\times10^{-4}$ and $\sigma_{\mathrm{stat}}(C_3/C_2)\simeq6.2\times10^{-3}$. At $N_{\mathrm{evt}}\simeq3.9\times10^{9}$, the combined correction gives $\sigma_{\mathrm{stat}}(C_2/C_1)\simeq3.1\times10^{-4}$ and $\sigma_{\mathrm{stat}}(C_3/C_2)\simeq3.8\times10^{-3}$. These minimum-bias samples correspond to approximately $1.1\times10^{7}$ and $2.0\times10^{8}$ selected 0\%--5\% events, respectively.

\subsection{Current Limitations and Further Tests}

The efficiency correction assumes a binomial detector response after conditioning on the single-track variables. However, the detector occupancy and TPC pad-row hits shared by tracks can induce correlations in finding a track and violate this assumption. The application on the finite detector acceptance, using the covered region to represent the uncovered region, imposes a stronger condition, because the tracks within and without the acceptance could be correlated. Additionally, the two-dimensional $\varepsilon_{\mathrm{acc}}(\eta,p_{\mathrm{T}})$ may not fully characterize the acceptance. For example, event-wise anisotropies and centrality-dependent $\phi$ distributions may alter higher-order cumulants beyond simple binomial sampling. Therefore, under the present fast-simulation conditions, the closure achieved by the corrections may not be concluded as a demonstration of the validity under realistic conditions. Nevertheless, the two correction strategies provide a practical technical basis for future studies.

At low collision energies, limited charged-particle multiplicity degrades the centrality resolution, while participant fluctuations, baryon-number conservation and residual correlations between $M_{\mathrm{ref}}$ and $N_{\mathrm{p}}$ remain after CBWC. Its performance should be tested for different CEE collision systems and beam energies, particularly for light-ion systems with lower multiplicities. The effect of volume fluctuations and centrality resolution is not included in this work, and it is better to have a test on complementary volume-fluctuation treatments.

This analysis assumes $p_{\mathrm{hit}}=0.6$ as the probability of a hit independently recorded by a TPC pad row. However, it is a basic parameter and necessary for the following analysis procedure, including the $\varepsilon_{\mathrm{rec}}$ estimation, effective acceptance coverage determination and $M_{\mathrm{ref}}$ calculation. How independently tracks are reconstructed also determines the validity of the binomial efficiency correction. Therefore, the full-simulation framework, which includes the full detector response, is required for the further analysis with a similar strategy. Furthermore, a centrality-dependent acceptance map can also be studied to estimate the validity range of the acceptance correction. It is also important to vary the $N_{\mathrm{hit}}^{\mathrm{cut}}$ value, the $\varepsilon_{\mathrm{rec}}$ threshold and the $n_{\sigma}^{\mathrm{p}}$ cut window. Understanding how the result changes could be valuable for the experimental systematics. The event requirements in \Cref{sec:statistical_precision} are extrapolated from the present sample and may change with realistic detector occupancy, centrality resolution, trigger and data-quality efficiencies and non-binomial response. These parts should be reproduced through the full-simulation framework.

\section{Summary}
Using $2\times10^{6}$ $^{\text{238}}$U+$^{\text{238}}$U collision events at $\sqrt{s_{\mathrm{NN}}}=$~2.1~GeV, corresponding to $\mu_{\mathrm{B}}\sim$~830~MeV, generated with JAM2 and processed through the CEE fast simulation, we develop an end-to-end proton-fluctuation analysis from detector-level proton selection and TPC-track-multiplicity centrality determination to the calculation of cumulants up to the fourth order ($C_1$--$C_4$) and their ratios ($C_2/C_1$, $C_3/C_2$ and $C_4/C_2$). Track reconstruction is simulated based on the TPC geometry, and the proton selection within dynamic cut windows ensures high purity $>95\%$. Within the TPC acceptance, the track-by-track binomial efficiency correction using the reconstruction efficiency ($\varepsilon_{\mathrm{rec}}(\eta,p_{\mathrm{T}},\phi)$) and the PID cut efficiency ($\varepsilon_{\mathrm{pid}}(y_{\mathrm{CM}},p_{\mathrm{T}})$) shows good closure against the true proton sample before random track loss. The $\phi$-averaged acceptance factor ($\varepsilon_{\mathrm{acc}}(\eta,p_{\mathrm{T}})$) is used to compensate statistically for the azimuthal gaps. The acceptance-corrected cumulants and cumulant ratios are consistent with the true values in a full $\phi$ coverage within statistical uncertainties. The statistical uncertainty of the cumulant ratios is also evaluated for the 0\%--5\% centrality class. Using $\sigma_{\mathrm{stat}}(C_4/C_2)\simeq0.086$ as a practical benchmark, the power-law extrapolation gives minimum-bias event requirements of approximately $2.1\times10^{8}$ for the efficiency-corrected result and $3.9\times10^{9}$ for the combined efficiency--acceptance correction. These event requirements are planning-scale projections and should be refined with larger simulated samples and realistic running conditions. This study, concluding both closure criteria and statistics guidelines, establishes an important basis for future CEE analyses and demonstrate the feasibility of proton-number fluctuation measurements.

\bibliography{references.bib}

\end{document}